\documentclass[12pt,oneside,english]{article}

\usepackage{amsmath}
 \usepackage{graphicx}
\usepackage[T1]{fontenc}
\usepackage[latin1]{inputenc}
\usepackage{babel}
\usepackage{setspace}
\usepackage{rotating}

\makeatletter

\begin{document}

\centerline{ \bf \large A possible quantum probability increase 
 of the cylindrical gravitational field }
\bigskip \bigskip

\centerline{\large \bf  D.   Bar}

\bigskip   \bigskip  \bigskip \bigskip \bigskip

\begin{abstract}
{\it As known, the  cylindrical gravitational    wave  have been canonically
quantized and its wave function, as the quantum one,  interpreted in probability
terms. We show in this work, using quantum Zeno methods, that this
probability may be substantially increased and even approach unity. For that we
first show, in detailed manner, that the cylindrical gravitational wave may be
discussed in the commutation number representation. We also discuss this field
in the transverse-traceless  (TT) gauge and calculate the related trapped 
surface. }

\end{abstract}

\bigskip \bigskip  \bigskip  \bigskip

\underline{Keywords}: Cylindrical gravitational wave, Quantization, Zeno effect

 \underline{Pacs numbers}: 03.70.+k, 04.30.-w, 03.65.Xp

   \bigskip

   \bigskip 
     
     \markright{INTRODUCTION} 
     
\protect \section{\bf INTRODUCTION}

The problem of quantizing the general gravitational wave (GW) \cite{note1} 
   \cite{thorne} have, 
theoretically,  been 
discussed  by different  persons and methods beginning from the earlier works of
Rosenfeld \cite{rosenfeld}, Bergmann \cite{bergmann} and Schwinger
\cite{schwinger1} to the canonical methods
of Dewitt \cite{dewitt},  Arnowitt-Desser-Misner (ADM) \cite{adm} and 
Dirac \cite{dirac}. Among the
supposed quantum characteristics of the   GW 
is the probability interpretation \cite{schiff,merzbacher} 
 which is extended from    the quantum regime  so that one     
   interprets \cite{kuchar1} the GW function as a probability
 amplitude \cite{schiff,merzbacher}.  
  This  problem of quantizing and 
 interpreting, in probability terms, 
  the GW  entails  in turn a second 
  interesting problem of how
 and in which way to increase the probabilty of some specific GW. 
 Taking the cylindrical GW as an example and assuming
 that one starts from a hypersurface with a cylindrical geometry 
  ane may ask how to increase its 
 probability so that this same cylindrical geometry may persist upon this
 hypersurface. \par 
 This problem is also found, under different terms and
 terminology, in other disciplines of general relativity.  
 Thus, a central issue of   quantum gravity \cite{kiefer} is the problem of  
 substantially 
  increasing  the  probability  of some  
 quantum  foam \cite{morris},  with a typical 
 order of magnitude 
 of the
 Planck-Wheeler length ($\frac{G\hbar}{c^3}$) \cite{mtw,hartle},  so that 
 it may be 
  realized in full grown 
 human scale with a unity probability (as phrased in \cite{thorn_0} 
 regarding   a Planck-Wheeler sized wormhole \cite{visser}).   \par  
  As known,  
 the mentioned first problem of quantizing the GW 
 finds its solution when one  restricts  the discussion to limited
dynamical regions of geometrodynamics such as the minisuperspace and
miniphasespace discussed by Dewitt \cite{dewitt}, Misner \cite{misner} and,
especially, by Kuchar \cite{kuchar1} which applied it for  quantizing
the cylindrical GW.  Kuchar did it by extending the extrinsic time idea,  which
was first introduced \cite{adm,peres,kuchar2} for the linearized theory
\cite{mtw,hartle} of general relativity,  to the nonlinear theory 
\cite{kuchar3} and,
especially, to the cylindrical spacetime.  This extrinsic time variable,  
which  is of the
 Tomonaga-Schwinger   many-fingered time kind
\cite{tomonaga,schwinger2},  is 
canonically conjugate to momentum and not to energy as is 
  the intrinsic time which serves more as a label to distinguish between
  spacelike hypersurfaces \cite{mtw} in a one-parameter family of them. \par The mentioned  property of being fully quantized makes the gravitational cylindrical minisuperspace a starting point for a possible general quantum gravity theory \cite{ashtekar,barbero,cho}. Moreover, it has even triggered a discussion \cite{hayward} about cylindrical blackholes and trapped cylinders. \par
   Using the  extrinsic time  variable 
one may obtain  \cite{kuchar1}, as  known from \cite{adm}, a formalism which is
identical to that used for discussing the parametrized \cite{kuchar1}
cylindrical massless scalar field in a flat Minkowskian background. This 
suggests, as
noted in \cite{kuchar1}, that all the results obtained from the later case may,
theoretically, 
be  applied also for the cylindrical GW   in a curved 
spacetime. These applications include also the canonical  quantization of the
cylindrical GW  and the derivation \cite{kuchar1} 
of a functional
time-independent Schroedinger-type equation for it. 
 Moreover, Kuchar has succeeded  \cite{kuchar1} 
to
apply for the cylindrical GW not only  quantum ideas 
 such as the 
  inner product
of states  \cite{schiff,merzbacher} or path independence of the evolution of 
them but also their 
probability interpretation \cite{schiff,merzbacher}. \par
In this work we use this successful quantization of the cylindrical
GW as a basis for discussing the mentioned second problem 
of finding the
conditions  under which the mentioned probability  increases and
even approaches unity.    In the quantum regime, where states change with both
time and space \cite{schiff,merzbacher},  
  there exists the known Zeno  effect \cite{zeno,aharonov,facchi,bar1} 
  which causes these changing  states to
  become constant and fixed and, therefore, to cause their probability, as in
   classical physics,  to become
  unity. This effect,  which were experimentally validated
  \cite{itano,kofman}, has three different versions:  \\
\noindent (1)  Repeating a large number of times, in
 a finite total time, the same experiment of checking  
  the present state of 
 some
 quantum system that has been prepared in an initial specific state so that in  
 the limit
 of a very large number of repetitions during the same total time 
 the initial state is preserved in time  \cite{zeno,itano}. \\ 
\noindent (2)  Performing many sets of experiments where each set is composed 
 of a large number of slightly different experiments  
 so that the overall effect is 
 as if one advances through     
  many  different possible paths of states (Feynman
 paths \cite{feynman,gert}). Thus,   taking  the limit of doing these 
 experiments 
  in a ``dense''  manner  one may ``realize''  \cite{aharonov,facchi}  any specific
 path of states in the sense that the probability to proceed along all its
 constituent states tends to unity. \\
 \noindent (3)  Simultaneously  performing   
 the same experiment 
  in a large number of non-overlapping spatial subregions  all included in 
  a finite total region  
  so that when
  these subregions  infinitesimally  shrink, keeping the total region fixed, 
   one avoid any spatial shifting of the state \cite{bar1}.  
  This may be
  explained by the example  of trying to locate a very small 
   particle-like 
  object in the finite total spatial region which become easier when this region
  is divided into several equal parts each occupying the same small object and
  the searching for it is done in each of these smaller regions. It is obvious  
  that the smaller become these regions in the finite total region 
   the probability to locate the small object in each of them grows
  so that in the limit in which they infinitesimaly shrink this probability
  tends to unity. \\
 \noindent  The first two cases (1)-(2)  are, respectively,  known
 \cite{aharonov,facchi} as 
 the static and 
    dynamic  Zeno  
 effects    while   the third (3) is  
   the space Zeno effect
 \cite{bar1}. \par  
   We  note  that whereas the  
quantum states  are related to  the ordinary  intrinsic time 
  the gravitational
cylindrical states  are characterized  by   \cite{kuchar1} an 
{\it extrinsic time} variable 
which is related (and actually borrows \cite{kuchar1}  its name) to the 
extrinsic
curvature. 
That is,  the 
gravitational state, represented by the probability  that the related
hypersurface has cylindrical geometry, changes  
 in spacetime with respect 
to {\it extrinsic time}. 
Thus, since  this extrinsic time, like the ordinary spatial variable, 
 is (see Sec VII in \cite{kuchar1}) canonically conjugate to
momentum (which is connected to extrinsic curvature) 
the corresponding gravitational Zeno effect should also have spatial
characteristics as in the mentioned space Zeno effect. \par
We directly show in this work, using space Zeno terms \cite{bar1}, that one may 
avoid any space
shifting of the cylindrical GW  thereby fixing its cylindrical
geometry and causing  its probability to approach unity. 
In the following we precede  this Zeno demonstration with a discussion 
wich  shows   that the  cylindrical  GW may be represented as  
a large number
of constituent parts in some finite region of space so that it may be discussed
in spatial Zeno terms \cite{bar1}.  The appropriate representation which 
enables one to do
so is the occupation number one
\cite{schiff,merzbacher}.  Thus, we discuss 
here,   in
detail,   this representation  
\cite{schiff,merzbacher}  in relation to  the cylindrical GW. 
 We note in this respect that although this
representation in the context of  cylindrical
GW  is mentioned and calculated in the literature (under the name of Fock quantization see \cite{kuchar1,ashtekar,barbero,macrina}) 
  we proceed here in a  detailed manner  along the methods used in \cite{schiff} for the harmonic oscillator.  \par 
As known from quantum mechanics in the occupation number representation 
\cite{schiff,merzbacher,mahan},
one may, theoretically,  prepare  any quantum state by merely 
applying the relevant creation and destruction operators
\cite{schiff,merzbacher}  any required 
number of times upon some
initial basis state.    
 One also  knows  from the canonical formalism \cite{adm,mtw,hartle} of general
relativity that it is possible to theoretically prepare  the geometry of some spacetime 
hypersurface by controlling
the form of the lapse and shift functions \cite{adm,mtw,hartle}. That is, any
specific theoretical 
evolution of spacetime should be preceded  by  determining beforehand 
these functions so that one can be sure (with a unity probability) that the
related spacetime is developed along the specified route.  This  
operation, characterized by 
 the determination of the lapse and shift functions,  should be related,
when discussing the quantum properties of the GW, 
to the corresponding  operation of the mentioned quantum creation and
destruction operators upon some initial basis state. We derive here in detail
the appropriate expressions which accordingly relate the cylindrical lapse 
and shift
functions to the creation and destruction operators. \par 
When the mentioned probability of the cylindrical GW approach unity (also in
consequence of the spatial Zeno effect)   the related GW   
produce certain effects upon the neighbouring spacetime through which it
proceeds such as implanting its cylindrical geometry upon it  and 
 giving  rise to some trapped surface \cite{eppley,abrahams,bar2,bar3}. We  
 follow the
developement of this cylindrical GW, once its probabilistic chances were greatly
increased,  and find its properties in the transverse-traceless (TT) gauge 
and also calculate the geometry of the generated  trapped
surface \cite{eppley,abrahams,bar2,bar3}.  \par    
In Section II we introduce the principal expressions \cite{kuchar1} related to the
Einstein-Rosen cylindrical
GW \cite{einstein} as, especially, represented in \cite{kuchar1,macrina}. 
In Section III we discuss the cylindrical 
GW  in the 
commutation number representation \cite{schiff,merzbacher} so that it may be
thought of as composed of a very large number of gravitational quanta which, 
 like the quantum ones  \cite{schiff,merzbacher}, are
 created
and destroyed by the  corresponding gravitational creation and destruction 
operators. In Section IV we relate the cylindrical lapse and shift functions
\cite{adm,mtw,hartle}  to the mentioned gravitational 
creation and destruction  operators. Note that the lapse and shift functions 
  participate (as emphasized in \cite{mtw} (see Sections 21.8 and 21.9 
  there))
in the determination of 
 the geometry of
spacetime hypersurfaces just as     the mentioned gravitational 
creation and destruction  operators  determine this geometry
through creating or (and) destroying the quantum 
components of the generating GW. 
 We note that the cylindrical lapse and shift functions were related in
\cite{kuchar1} to the Einstein-Rosen parameters 
(see Eqs (29)-(30) in \cite{kuchar1}).  The detailed calculation relating the
cylindrical lapse and shift functions to the creation and destruction  
operators is shown in
Appendix $A$. In Section V we  show that, beginning
from a gravitational cylindrical geometry in some hypersurface, the 
probability to find the
same geometry upon this hypersurface tends to unity   in the limit
of the space Zeno effect \cite{bar1}. As mentioned,  we rather discuss the 
space
Zeno effect   and not the (intrinsic) time analogue of it because in the 
cylindrical
geometry one discusses \cite{kuchar1} the extrinsic time variable  which 
 is canonicaly conjugate to
momentum (Sec VII in \cite{kuchar1}) just as is any spatial variable. We note 
that it has 
been shown \cite{bar4}, 
 using the examples of the quantum bubble and open-oyster processes
\cite{mahan,mattuck}, that the mentioned static  and dynamic quantum Zeno 
effects 
are also valid in quantum field theory \cite{mahan,mattuck}. We also note in this
respect that the quantum Zeno effect were discussed \cite{camacho} in the
framework of gravitomagnetism \cite{mashhoon}. The detailed calculations of the
appropriate probability is shown in Appendix $B$. In Section VI we 
graphically corroborate our
theoretical results  so that one may
see how the probability approach unity in the Zeno limit. In
Section VII we discuss the cylindrical GW in the transverse-traceless (TT) 
gauge
which is characterized by a very simplified formalism \cite{mtw} in which, for 
example, the
number of independent components of the related GW is minimal \cite{mtw}. 
In Section VIII we discuss, using the method in \cite{eppley},  the related 
embedded trapped surface \cite{mtw,eppley,abrahams,bar2,bar3} resulting from the
passing cylindrical  GW. In Section IX we summarize our discussion in a
Concluding Remarks Section.

\markright{THE EINSTEIN-ROSEN CYLINDRICAL GRAVITATIONAL WAVE}

\protect \section{ The Einstein-Rosen Cylindrical gravitational wave}

 A spacetime is considered to be cylindrically symmetric \cite{kuchar1} 
if and only if one can show that there sxists 
 a coordinate system $(t, \ r, \ \phi, \ z)$,  $ -\infty <  t < +\infty$, 
 $\infty > r \geq 0$, $ 2\pi > \phi \geq 0$, $ -\infty < z <+\infty$ in which the line
 element becomes 
 \begin{equation}  \label{e1} ds^2 =-(N^2 - e^{(\psi- \gamma)}N_1^2)dt^2+2N_1dtdr+
 e^{(\gamma- \psi)}dr^2+R^2e^{-\gamma}d\phi^2+e^{\gamma}dz^2,   \end{equation}
where $ R \geq 0$ and $\gamma, \ \psi, \ N, \ N_1$ are functions of $t$ and $r$.
The former dependence of the nonzero metric tensor components $g_{11}, g_{22},
g_{33}, g_{00}, g_{01}$ upon the functions  $\gamma, \psi, R, N, N_1$ is, 
especially, designed \cite{kuchar1} to suit the  ADM \cite{adm} canonical formulation of general
relativity. Thus, the $N$ and $N_1$ are,  respectively, the known ADM lapse and
radial shift functions \cite{kuchar1,mtw}. The coordinates $\phi$ and $z$ are, 
essentially,  fixed except for a possible trivial transformation of
$\bar{\phi}=\pm \phi+\phi_0$ and $\bar{z}=az+z_0$ whereas $t$ and $r$ may be
subject, without changing the form of the line element from (\ref{e1}), 
 to the more general transformation 
 \begin{equation}\label{e2} \bar{t}=\bar{t}(t,r), \ \ 
\bar{r}=\bar{r}(t,r)  \end{equation}
One may show \cite{kuchar1}, using Killing vectors in the $(t, \ r, \
\phi, \ z)$ system,  that the functions $R$ and $\gamma$ are scalars. 
Thus,  since the metric tensor coefficients depends, as mentioned, only upon $t$
and $r$ one may write \cite{kuchar1} the $(t, r)$ part of the line element
(\ref{e1}) in the following conformally flat form 
\begin{equation} \label{e3} ds^2=e^{(\bar{
\gamma}-\psi)}(-d\bar{t}^2+d\bar{r}^2)+R^2e^{-\psi}d\phi^2+e^{\psi}dz^2, 
\end{equation}
where the  $R$, $\psi$,  $\phi$ and $z$ are not barred due to their
mentioned essential invariancy. Now, as emphasized in \cite{kuchar1}, if one
writes the Einstein field equations for the line element (\ref{e3}) one may
realize that $R$ must be a harmonic function which satisfies $\frac{\partial^2
R}{\partial \bar{t}^2}- \frac{\partial^2
R}{\partial \bar{r}^2}=0$. Thus, one may assume \cite{kuchar1} $R$ to be a new radial
coordinate and $T$ the time coordinate corresponding to it. That is,  as 
emphasized in 
\cite{kuchar1},
the Einstein-Rosen coordinates can  be uniquely and rigorously defined by
invariant prescriptions so that the line element (\ref{e3}) may be written as 
 \begin{equation} \label{e4} ds^2=e^{(
\Gamma-\psi)}(-dT^2+dR^2)+R^2e^{-\psi}d\phi^2+e^{\psi}dz^2 
\end{equation} 
In such case the Einstein vacuum equations are considerably simplified and
reduce to the following three equations 
\begin{equation} \label{e5} \frac{\partial^2
\psi}{\partial T^2}- \frac{\partial^2
\psi}{\partial R^2} -R^{-1}\frac{\partial
\psi}{\partial R}=0  \end{equation}

\begin{equation} \label{e6} \frac{\partial \Gamma}{\partial R}=
\frac{1}{2}R\biggl((\frac{\partial
\psi}{\partial T})^2 +  (\frac{\partial
\psi}{\partial R})^2\biggr)
   \end{equation}
   
 \begin{equation} \label{e7} \frac{\partial \Gamma}{\partial T}=
 R\frac{\partial
\psi}{\partial T} \frac{\partial
\psi}{\partial R} 
   \end{equation}
   
  As emphasized in \cite{kuchar1}, Eq (\ref{e5}) has exactly the same form as the
  wave equation of the cylindrically symmetric massless scalar field $\psi$
  advancing in a Minkowskian spacetime whereas Eqs (\ref{e6}) and (\ref{e7})
  are, respectively, the energy density and the radial momentum density of this
  field in cylindrical coordinates. The solution of Eq (\ref{e5}) is obtained by
  using the separation of variables method \cite{macrina} so that the resulting
  wave function for a particular wave number $k$ is 
  \begin{equation}  \label{e8} \psi_k(R,T)=
J_0(kR)\bigl(A(k)e^{(ikT)}+A^*(k)e^{-(ikT)}\bigr),  
\end{equation}  
where $j_0(kR)$ is the bessel function of order zero \cite{abramowitz} and
$A(k), \ A^*(k)$ are the amplitude and its complex conjugate of the solution to
the time part of Eq (\ref{e5}). Note that here we assume, as generally done in
the relevant literature,  that $c=\hbar=1$ so that $w=k=p$ where $w, \ k, \ p$ are
respectively the frequency, wave number and momentum of some mode. Since 
 $k$    is a continuous parameter one may obtain
the general solution to Eq (\ref{e5}) by integrating over all the modes $k$. 
Thus, the relevant general wave function is 
\begin{equation}  \label{e9}  \psi(R,T)=\int_0^{\infty} dk
J_0(kR)\bigl(A(k)e^{(ikT)}+A^*(k)e^{-(ikT)}\bigr)  
\end{equation}
The canonical conjugate momentum $\pi_{\psi}(T,R)$ may be obtained 
\cite{macrina} by using the
Hamilton equation \begin{equation}  \label{e10}  \frac{\partial \psi}{\partial
t}=\{\psi,H\}, \end{equation}  where $\psi$ is given by Eq (\ref{e9}), the curly
brackets at the right denote the Poisson brackets  and the
Hamilton function $H$ is \begin{equation} \label{e11}  H=\int_0^{\infty}
dr\biggl(\tilde{N}\tilde{H}+\tilde{N^1}\tilde{H}_1\biggr)  \end{equation} 
The quantities $\tilde{H}$ and $\tilde{H}_1$ are respectively the rescaled 
superHamiltonian and supermomentum which where shown in \cite{kuchar1} (see Eqs
(93)-(97) and (106)-(108) in \cite{kuchar1}) to be 
\begin{eqnarray} && \tilde{H}=R_{,r}\Pi_T+T_{,r}\Pi_R+
\frac{1}{2}R^{-1}\pi_{\psi}^2+\frac{1}{2}R\psi_{,r}^2 \label{e12} \\
&& \tilde{H}_1=T_{,r}\Pi_T+R_{,r}\Pi_R+\psi_{,r}\pi_{\psi},  \nonumber 
\end{eqnarray}
where the suffixed apostroph denote differentiation with respect to $r$ and $\Pi_T, \
\Pi_R$ are the respective momenta canonically conjugate to $T$ and $R$. The
quantities $\tilde{N}$  and $\tilde{N}_1$ respectively denote the rescaled 
lapse and shift
function $N$, $N_1$ (see Eq (96) in \cite{kuchar1}). Thus, $\pi_{\psi}(T,R)$ 
were shown \cite{macrina} to have
the form 
\begin{eqnarray}  &&  \pi_{\psi}(T,R)=iRR_{,r}\int_0^{\infty} dk k
J_0(kR)\bigl(A(k)e^{(ikT)}-A^*(k)e^{-(ikT)}\bigr) \label{e13} \\ &&
-RT_{,r}\int_0^{\infty} dk kJ_1(kR)\bigl(A(k)e^{(ikT)}+A^*(k)e^{-(ikT)}\bigr), 
\nonumber 
\end{eqnarray}
where $J_1(kR)$ is the first order Bessel function \cite{abramowitz} which may
be obtained by differentiating $J_0(kR)$ with respect to $R$ as
$J_0(kR)_{,R}=-kJ_1(kR)$. The initial data for $\psi(T,R)$ and $\pi_{\psi}(T,R)$
are calculated for $T=0$ and $R=r$ and are, respectively, denoted by $Q(r)$ and
$P(r)$  as follows
\begin{equation}   Q(r)=\psi_0(r)=\psi(T,R)|_{T=0,R=r}=\int_0^{\infty} dk
J_0(kr)\bigl(A(k)+A^*(k)\bigr)  \label{e14} 
\end{equation}
\begin{equation}   P(r)=\pi_{\psi_0}(r)=\pi_{\psi}(T,R)|_{T=0,R=r}=
 ir\int_0^{\infty} dk
kJ_0(kr)\bigl(A(k)-A^*(k)\bigr) \label{e15}  
\end{equation}
Solving  the last two equations for $A(k)$ and $A^*(k)$ one   obtains
\begin{equation}   A(k)=\frac{1}{2}\int_0^{\infty} dr
J_0(kr)\bigl(krQ(r)-iP(r)\bigr)  \label{e16}
\end{equation}
\begin{equation}   A^*(k)=\frac{1}{2}\int_0^{\infty} dr
J_0(kr)\bigl(krQ(r)+iP(r)\bigr)  \label{e17}
\end{equation}
One may show, using Eqs (\ref{e16})-(\ref{e17}), that the variables $A(k),\
A^*(k)$ satisfy the following 
Poisson brackets \begin{eqnarray} && \{A(k),A^*(k')\}
=\int_0^{\infty}dr\biggl[\frac{\delta(A(k))}{\delta(Q(r))}
\frac{\delta(A^*(k'))}{\delta(P(r))}-\frac{\delta(A(k))}{\delta(P(r))}
\frac{\delta(A^*(k))}{\delta(Q(r))}\biggr]=
\frac{i\delta(k-k')}{2}  \nonumber \\ &&  
\{A(k),A(k')\}=\{A^*(k),A^*(k')\}=0,    
 \label{e18}  \end{eqnarray}
 where  use was made of the relation  
   \cite{macrina} $$ \int_0^{\infty}dr r
J_n(kr)J_n(k'r)=\frac{1}{k'}\delta(k-k'), \ \ \ n=0, 1, 2, 3....$$ 
In a similar manner, using Eqs (\ref{e14})-(\ref{e15}), it is possible to show  
that the variables 
$Q(r)$ and $P(r)$ satisfy the following 
Poisson brackets
\begin{eqnarray} &&   
\{Q(r),P(r')\}
=\int_0^{\infty}dk\biggl[\frac{\delta(Q(r))}{\delta(A(k))}
\frac{\delta(P(r'))}{\delta(A^*(k))}-\frac{\delta(Q(r))}{\delta(A^*(k))}
\frac{\delta(P(r'))}{\delta(A(k))}\biggr]=
-2i\delta(r-r') \nonumber \\ 
&&  \{Q(r),Q(r')\} =\{P(r),P(r')\}  =0,
\label{e19} 
\end{eqnarray}
where use was made of the  
relation  \cite{macrina} 
$$\int_0^{\infty}dkkJ_n(kr)J_n(kr')=\frac{\delta(r-r')}{r'}, \ \ \
 n=0, \ 1, 2, \ldots$$
Using Eqs (\ref{e6}), (\ref{e9}) and the unnumbered relation written just after Eq 
(\ref{e18})  
one may obtain the following expression for the energy  $\Gamma$ 
\cite{macrina} 
\begin{equation}   \Gamma=2\int_0^{\infty} dk
kA(k)A^*(k)   \label{e20} 
\end{equation}
As shown in \cite{macrina} the former expression (\ref{e20}) is for $R$ at
infinity and it shows that the energy remains finite for this case. We are
interested in other expression for the energy in terms of $Q(r)$ and $P(r)$
which is derived from the expression (\ref{e20}) 
by using Eqs (\ref{e16})-(\ref{e17}) and  the Poisson brackets (\ref{e19}) 
\begin{equation}   \Gamma-k=\int_0^{\infty}dr\biggl(\frac{P^2(r)}{2r}+
\frac{rk^2Q^2(r)}{2}\biggr),  \label{e21}  \end{equation} 
where the $k$ term is obtained by using the  relation (\ref{e19}) 
and 
  the unnumbered relation   just after it. The  term under the integral sign 
   involving $P^2(r)$ and
  $Q^2(r)$  is obtained by using   the following orthogonality expression 
  which is valid for
any $n= 0, \ 1, \ 2,\ \ldots$
\begin{equation}   \label{e22}
\int_0^{\infty}dr'r'\int_0^{\infty}dkkJ_n(kr)J_n(kr')f(r')=f(r) 
\end{equation} 
Note that the energy at the right hand side of Eq (\ref{e21}) corresponds to the
energy of the harmonic oscillator \cite{schiff} $H_{harmonic \
oscilltor}=\frac{P^2}{2m}+\frac{1}{2}mw^2r^2$ not only in its form but also in
its role in the following commutation relations (\ref{e23})-(\ref{e24}) and in 
Eqs  (\ref{e25})-(\ref{e32}).

\markright{THE OCCUPATION NUMBER REPRESENTATION FOR CYLIDRICAL....}

\protect \section{The occupation number representation for  cylindrical field}

 In this section we apply the occupation number formalism
 \cite{schiff,merzbacher} for the cylindrical
 GW. As known \cite{schiff,merzbacher}, the passage from the
 classical domain to the quantum one entails regarding the classical variables
 as operators \cite{schiff,merzbacher}  and the  change,  for any
 function $f$,  of  $f^* \rightarrow f^+$ where $f^+$ is the hermitian adjoint of
 $f$ \cite{schiff,merzbacher}. Also, the known Poisson brackets $\{f,f^*\}$ are 
 replaced by the quantum
 commutation ones $[f,f^+]$ obtained through $\{f,f^*\} \rightarrow \frac{[f,f^+]}{i\hbar}$ where, as
 mentioned, we assign the unity value to $\hbar$.  In the following  
 we use Eqs
 (\ref{e18}) for calculating   
 the commutation relations 
 between $Q(r), \ P(r)$  
  from Eq (\ref{e14})-(\ref{e15})  and the  energy   
    $\int_0^{\infty}dr\biggl(\frac{P^2(r)}{2r}+
\frac{rk^2Q^2(r)}{2}\biggr)$  from  Eq
 (\ref{e21}).     Thus,   the commutation 
relation between the
 observable $Q(r)$ from Eq (\ref{e14}) and
 the former energy $(\Gamma-k)$ from Eq (\ref{e21})  is
 \begin{eqnarray} && [Q(r),(\Gamma-k)]  = \biggl[ \int_0^{\infty} dk
J_0(kr)\biggl(A(k)+A^+(k)\biggr),2\int_0^{\infty} dk'
k'A(k')A^+(k')-k \biggr]= \nonumber \\  
&& =\int_0^{\infty}\int_0^{\infty}dkdk'J_0(kr)k'\biggl(A(k')
\delta(k-k')-A^+(k')\delta(k'-k)\biggr)=   \label{e23} \\
&&=\int_0^{\infty}dk'k'J_0(k'r)\biggl(A(k')
-A^+(k')\biggr)=\frac{P(r)}{ir} \nonumber \end{eqnarray}
Likewise, using  again Eqs (\ref{e18})  and the expression   
(\ref{e21}) for the
relevant energy $(\Gamma-k)$,  we  calculate the 
commutation relation
 between $P(r)$ from Eq (\ref{e15}) and this energy   
  \begin{align}& [P(r),(\Gamma-k)]=\biggl[ir \int_0^{\infty} dk k
J_0(kr)\biggl(A(k)-A^+(k)\biggr),2\int_0^{\infty} dk'
k'A(k')A^+(k')-k\biggr]= \nonumber \\  &= 
-\frac{r}{i}\int_0^{\infty}dkk^2J_0(kr)\biggl(A(k)
+A^+(k)\biggr)= -\frac{rk^2Q(r)}{i} \label{e24}  \end{align}   
Using the Dirac's ket and bra notation \cite{schiff,merzbacher} 
for the matrix representation of the obsrvables $Q(r)$, $P(r)$,   
$(\Gamma-k)$ and taking into account that $|j\!><\!j|$ is unit operator 
one may obtain from the two ends of the first commutation relation (\ref{e23})
\begin{eqnarray} && <\!k|Q|j\!><\!j|(\Gamma-k)|l\!>-
<\!k|(\Gamma-k)|j\!><\!j|Q|l\!>= \label{e25} \\ && = 
(E_l-E_k)<\!k|Q|l\!>=
\frac{<\!k|P|l\!>}{ir},   \nonumber  \end{eqnarray}
where $E_l$ and $E_k$ are the  eigenvalues of the energy operator $(\Gamma-k)$
which, respectively, correspond to the kets $|l\!>$ and $|k\!>$.   
 In a similar manner one may obtain  from the two ends of the 
 second commutation relation (\ref{e24}) 
\begin{eqnarray}  && <\!k|P|j\!><\!j|(\Gamma-k)|l\!>-
<\!k|(\Gamma-k)|j\!><\!j|P|l\!>= \label{e26} \\ && = (E_l-E_k)<\!k|P|l\!>
=ik^2r<\!k|Q|l\!>  \nonumber  \end{eqnarray}  
Solving  the last equation for $<\!k|Q|l\!>$ and substituting in (\ref{e25}) 
one obtains for $(E_l-E_k)$ \begin{equation} (E_l-E_k)=\pm k 
\label{e27} \end{equation} 
Now, multiplying Eq (\ref{e25}) by $-irk$ and adding to (\ref{e26}) 
one obtains
\begin{align} & (E_l-E_k-k)<\!k|P|l\!>+irk\biggl(E_l-E_k-k\biggr)<\!k|Q|l\!>= 
 \label{e28} \\ &
(E_l-E_k-k)<\!k|(P+irkQ|l\!>=0 \nonumber  \end{align}  
From the last equation one may realize that $<\!k|(P+irkQ)|l\!>$ is 
 different from
zero only when $E_k=E_l-k$. That is, operating  with the operator 
$(P+irkQ)$ on the ket $|l\!>$ results in some multiple of the ket $|k\!>$ with 
an
energy lower by $k$ than that of the ket $|l\!>$. If, on the other hand,  
   Eq (\ref{e25}) is multiplied by $irk$ and then added to (\ref{e26}) 
   one comes with the result that operating with the hermitian adjoint 
   operator $(P-irkQ)$  on the ket $|l\!>$ results in some multiple of the ket 
   $|k\!>$ with 
an
energy higher by $k$ than that of the ket $|l\!>$.  Note that for the harmonic
oscillator these two operators change roles so that $(P+irkQ)$ is the energy
raising operator whereas $(P-irkQ)$ is the energy lowering one (see Section 25
in \cite{schiff}). Now, since the energy  must
be positive  one can not apply indefinitely the lowering operator $(P+irkQ)$ on
any ket unless there exists a lowest energy eigenstate $|0\!>$ so that 
$(P+irkQ)|0\!>=0$ (compare with the  analogous discussion in \cite{schiff}
regarding the harmonic oscillator). The lowest energy eigenvalue corresponding
to the lowest eigenstate may be found by first operating with  $(P-irkQ)$ on 
$(P+irkQ)|0\!>$ and using the quantum version of the  relations (\ref{e19})      
\begin{align} &
(P-irkQ)(P+irkQ)|0\!>=
\biggl(P^2+r^2k^2Q^2- 
irk(QP-PQ)\biggr)|0\!>=    \label{e29} \\ &=
2r\biggl(\frac{P^2}{2r}+\frac{rk^2Q^2(r)}
{2}-k\delta(r-r')\biggr)|0\!>=0  \nonumber \end{align}
We, now, divide the last expression by $2r$ and integrate the obtained 
result with
respect to $r$    
  \begin{align} & \biggl\{\int_0^{\infty}dr\biggl(\frac{P^2}{2r}+\frac{rk^2Q^2(r)}
{2}-k\delta(r-r)\biggr)\biggr\}|0\!>= 
\biggl\{\int_0^{\infty}dr\biggl(\frac{P^2}{2r}+\frac{rk^2Q^2(r)}
{2}\biggr)-   \label{e30} \\ & -  k\biggr\}|0\!>=
 0  \nonumber
\end{align}
We, thus, see that the eigenvalue which corresponds to the lowest energy eigenstate
$|0\!>$ is  $k$.    
Note that in the harmonic oscillator case one obtains a value of $\frac{k}{2}$
for the lowest eigenvalue (see Section 25 in \cite{schiff}) because the value
obtained there for the commutation relation $[Q,P]$ is half the value obtained 
here.  As mentioned, application of
the raising operator $(P-irkQ)$ on an arbitrary ket results in raising its
energy by $k$ and repeated application of it generates a 
sequence of
eigenstates that may appropriately be denoted by $|n\!>$. The energy eigenvalues
of this sequence may be expressed by 
\begin{equation} E_n=(n+1)k, \ \ \ n=0, \ 1, \ 2, \ 3, \ldots 
 \label{e31} \end{equation}
Now, if we, respectively,  multiply the lowering and raising operators 
$(P+irkQ)$,  $(P-irkQ)$ by $-\frac{iJ_O(kr)}{2}$ and $\frac{iJ_0(kr)}{2}$ and 
integrating the
resulting expressions with respect to $r$ from $r=0$ to $r=\infty$ one,
actually,
obtains the variables $A(k)$ and $A^+(k)$ as seen from Eqs
(\ref{e16})-(\ref{e17}). Thus, one may call the variable $A(k)$, 
in analogy with
the harmonic oscillator case \cite{schiff,merzbacher}, the energy lowering or
destruction operator and the variable $A^+(k)$ may, correspondingly, 
be called the energy raising or creation operator. This may also be seen from
Eqs (\ref{e20})-(\ref{e21}) which may be  read 
 $$ 2\int_0^{\infty} dk
kA(k)A^+(k) =\int_0^{\infty}dr\biggl(\frac{P^2(r)}{2r}+
\frac{rk^2Q^2(r)}{2}\biggr)+k, $$ and also to read, when the 
places of $A(k)$ and
$A^+(k)$ are commuted, 
  $$ 2\int_0^{\infty} dk
kA^+(k)A(k) =\int_0^{\infty}dr\biggl(\frac{P^2(r)}{2r}+
\frac{rk^2Q^2(r)}{2}\biggr)-k $$ 
The last equation may be written as  
\begin{equation} \biggl(2\int_0^{\infty} dk
A^+(k)A(k)+1\biggr)k =\int_0^{\infty}dr\biggl(\frac{P^2(r)}{2r}+
\frac{rk^2Q^2(r)}{2}\biggr) \label{e32} \end{equation}
Comparing Eqs (\ref{e31}) and (\ref{e32}) one may realize, as done in the
corresponding harmonic oscillator case \cite{schiff}, that to the eigenvalues
$n$ of Eq (\ref{e31}) there corresponds the  operator $N$ 
\begin{equation} N= \int_0^{\infty}dkN_k = \int_0^{\infty} dk
(2A^+(k)A(k)),    \label{e33} \end{equation}
so that $N_k=2A^+(k)A(k)$.  Now, as realized from all the former equations,  the
variable $k$ was considered to be  continuous but one may refer to it as a
discrete parameter (see, for example,  the discussion in P. 502 at
\cite{schiff}  and  Eqs (54.1)  and (55.11) there) in which case one obtains the same former expressions and 
results except for replacing all the integrals over $k$  by
corresponding summations.  In such case Eq (\ref{e33}) may be written as 
$N=\sum_kN_k=\sum_k2A^+_kA_k$ where $N_k=2A^+_kA_k$. 
Thus, it may be shown, as for the corresponding 
harmonic
oscillator case, that considering the representation in which each $N_k$ is
diagonalized,  the states of the quantized field may be represented, as done in
\cite{schiff},  by the kets
$|n_1, n_2, n_3, \ldots, n_k, \ldots \!>$ in which each $n_k$ is a positive
integer or zero and it is an eigenvalue of $N_k$. Also, since in this
representation only the diagonal matrix elements of $A^+(k)A(k)$ are nonzero one
may write such an element as 
\begin{equation} <\!n_k|A^+(k)A(k)|n_k\!>=
<\!n_k|A^+(k)|n_k'\!><\!n_k'|A(k)|n_k\!>=|\lambda_{n_k}|^2=n, 
 \label{e34} \end{equation}
where a summation over a complete set $n_k'$ is meant and $\lambda_{n_k}$ is
equal to $n^{\frac{1}{2}}$. Thus, from the last equation one may obtain the
relations 
\begin{equation} A(k)|n_1, n_2, n_3, \ldots, n_k, \ldots \!>=
n_k^{\frac{1}{2}}|n_1, n_2, n_3, \ldots, n_k-1, \ldots \!> 
 \label{e35} \end{equation}
\begin{equation} A^+(k)|n_1, n_2, n_3, \ldots, n_k, \ldots \!>=
(n_k+1)^{\frac{1}{2}}|n_1, n_2, n_3, \ldots, n_k+1, \ldots \!> 
 \label{e36} \end{equation}
The last two equations shows that  $A(k)$ and $A^+(k)$ are, respectively, 
the destruction and creation operators for the state $k$ of the field.
We, thus, have shown that the quantized cylindrical GW may be 
discussed in the
commutation number representation in which it may be thought of as composed 
of a
 large number of particle-like components inhabiting a certain finite spacetime 
 region. This special representation enables us, as mentioned,  to discuss the
 cylindrical GW in space Zeno terminology as we do in the following
 Sections V-VI.

\markright{THE LAPSE AND SHIFT FUNCTIONS CORRESPONDING TO THE....}

\protect \section{The lapse and shift function corresponding to the cylindrical
gravitational wave}

For finding the appropriate lapse and shift functions $N^{\alpha}$ which
correspond to the cylindrical GW's in the canonical formalism we begin from the
Hamilton equation (\ref{e10}) in its quantum version 
 $ \dot{\psi} = \frac{1}{i}[\psi, \ H ]$
where $\psi$ is given by Eq (\ref{e9}) and the square brackets  at the right  
denote the quantum 
 commutation relations. The expression for the energy $H$ which involves 
the operators
$A(k)$ and $A^+(k)$  is given by Eqs (\ref{e20})-(\ref{e21})  so substituting 
from Eqs
(\ref{e9}) and (\ref{e20})-(\ref{e21})  in   $ \dot{\psi} = \frac{1}{i}[\psi, \ H ]$ 
one obtains \begin{eqnarray} &&
\frac{\partial \biggl(\int_0^{\infty} dk
J_0(kR)\bigl(A(k)e^{(ikT)}+A^+(k)e^{-(ikT)}\bigr)\biggr)}{\partial t} =
\label{e37} \\ && =
\frac{1}{i}\biggl[\int_0^{\infty} dk
J_0(kR)\bigl(A(k)e^{(ikT)}+A^+(k)e^{-(ikT)}\bigr), 2\int_0^{\infty} dk
kA(k)A^+(k) -k \biggr] \nonumber 
\end{eqnarray}  
In the following we introduce the lapse and shift functions (see, especially, Eqs (\ref{e38})-(\ref{e39})) using the method described in \cite{macrina}. Thus, 
the time differentiation at the left hand side of the last equation may be
performed by using the embedding \cite{macrina} 
$X^{\alpha}=(T(r), R(r), \phi, z)$ which maps
the hypersurface $\Sigma$ at $t=constant$ into the flat spacetime as $X^{\alpha}
: \Sigma \rightarrow  R^4$. The embeddings $X^{\alpha}=(T(r), R(r), \phi, z)$
are cylindrically symmetric slices in spacetime where, due to this symmetry, 
 $T(r)$ and $R(r)$ do not depend on $\phi$ and  $z$. Thus, the time
 differentiation $\frac{\partial \psi(r)}{\partial t}$ may be written as 
   \begin{equation}  \frac{\partial \psi(r)}{\partial t}=
   \frac{\partial \psi(r)}{\partial X^{\alpha}(r)}\frac{dX^{\alpha}(r)}{dt}=
   N^{\alpha}\frac{\partial \psi(r)}{\partial X^{\alpha}(r)}, \label{e38} 
   \end{equation}
where \cite{macrina} $\frac{dX^{\alpha}(r)}{dt}=N^{\alpha}$. Thus, 
using the last equation, the
relation \cite{abramowitz} $\frac{dJ_0(r)}{r}=-J_1(r)$ and
the fact that $\psi$ does not depend on $\phi$ and $z$ one
may calculate the commutation relation from Eq (\ref{e37}) as 
\begin{align} &\frac{\partial \psi(r)}{\partial t}=
 N^{\alpha}\frac{\partial \psi(r)}{\partial X^{\alpha}(r)}=
N^0\frac{\partial \psi(r)}{\partial T(r)}+
N^1\frac{\partial \psi(r)}{\partial R(r)}=iN^0\int_0^{\infty} dk k
J_0(kR)\bigl(A(k)e^{(ikT)}- \nonumber \\ & -  A^+(k)e^{-(ikT)}\bigr) - 
N^1\int_0^{\infty} dk k
J_1(kR)\bigl(A(k)e^{(ikT)}+A^+(k)e^{-(ikT)}\bigr)=  \label{e39}  \\ & = 
 2\int_0^{\infty}dk
\int_0^{\infty}dk'J_0(kr)k'\biggl(e^{(ikT)}A(k')\frac{\delta(k-k')}{2}-
e^{-(ikT)}A^+(k')\frac{\delta(k'-k)}{2}\biggr)= \nonumber \\ & =
\int_0^{\infty}dkkJ_0(kr)\biggl(e^{(ikT)}A(k)-e^{-(ikT)}A^+(k)\biggr) 
\nonumber
\end{align}
  Since the last equation involves real and imaginary expressions we have to
decompose it into two equations one of which  relates  the real expressions 
among themselves and the second 
the imaginary ones.  Thus, the equation involving the real expressions is
\begin{align} & -N^0\int_0^{\infty} dk k
J_0(kR)\sin(kT)\bigl(A(k)+ A^*(k)\bigr) - 
N^1\int_0^{\infty} dk k
J_1(kR)\cos(kT) \nonumber \\ & \cdot \biggl(A(k)+ A^+(k)\biggr) =
 \int_0^{\infty} dk k
J_0(kR)\cos(kT)\biggl(A(k)-A^+(k)\biggr) \label{e40} 
\end{align}
And that involving the imaginary ones is
\begin{align} & N^0\int_0^{\infty} dk k
J_0(kR)\cos(kT)\bigl(A(k)- A^+(k)\bigr) - 
N^1\int_0^{\infty} dk k
J_1(kR)\sin(kT)\bigl(A(k)- \nonumber \\ & - A^+(k)\bigr) =
 \int_0^{\infty} dk k
J_0(kR)\sin(kT)\bigl(A(k)+A^+(k)\bigr) \label{e41} 
\end{align}
We integrate both sides of the last two equations (\ref{e40})-(\ref{e41}) 
over $r$ from $r=0$ to
$r=\infty$ and also for avoiding the intricacy of the resulting expressions and
facilitating the following calculation we label these expressions as
\begin{eqnarray} && C_1=\int_0^{\infty}dr \int_0^{\infty}dk
kJ_0(kr)\sin(kT)\bigl(A(k)+A^+(k)\bigr)  \nonumber \\
&& C_2=\int_0^{\infty}dr \int_0^{\infty}dk
kJ_1(kr)\cos(kT)\bigl(A(k)+A^+(k)\bigr)  \label{e42} \\
&& C_3=\int_0^{\infty}dr \int_0^{\infty}dk
kJ_0(kr)\cos(kT)\bigl(A(k)-A^+(k)\bigr)  \nonumber \\
&& C_4=\int_0^{\infty}dr \int_0^{\infty}dk
kJ_1(kr)\sin(kT)\bigl(A(k)-A^+(k)\bigr)  \nonumber 
\end{eqnarray} 
Thus, Eqs (\ref{e40})-(\ref{e41})  integrated over $r$, may be
compactly written as 
\begin{equation} -C_1N^0-C_2N^1=C_3, \ \ \  C_3N^0-C_4N^1=C_1  \label{e43} 
\end{equation}
Solving the last two equations for $N^0$ and $N^1$ one obtains 
\begin{equation} N^0=\frac{(C_1C_2-C_4C_3)}{(C_1C_4+C_2C_3)}, \ \ \ 
N^1=-\frac{(C_1C_1+C_3C_3)}{(C_1C_4+C_2C_3)}  \label{e44} 
\end{equation} 
The expressions  $(C_1C_4+C_2C_3)$, $(C_1C_2-C_4C_3)$,  $(C_1C_1+C_3C_3)$ are
calculated in Appendix $A$ and are, respectively,  given by Eqs 
(\ref{$A_2$})-(\ref{$A_4$})
there. Thus, 
substituting these results   of Appendix $A$ 
in Eqs (\ref{e44})
we obtain the following results for $N^0$ and $N^1$.
\begin{equation} N^0=\frac{\int_0^{\infty}dkk\sin(2Tk)\biggl(2A(k)A^+(k)-
\frac{1}{2}\biggr)}{\int_0^{\infty}dkk\biggl(A(k)A(k)-A^+(k)A^+(k)-\frac{1}{2}\biggr)
} \label{e45} 
\end{equation}

\begin{equation}
N^1=\frac{2\int_0^{\infty}dkk^2\biggl\{\cos(2Tk)\biggl(A(k)A(k)+A^+(k)A^+(k)\biggr)-\biggl(2A(k)A^+(k)
-\frac{1}{2}\biggr)\biggr\} }{r\int_0^{\infty}dkk\biggl(A(k)A(k)-A^+(k)A^+(k)-\frac{1}{2}\biggr)
}  \label{e46}
\end{equation}
The last Eqs (\ref{e45})-(\ref{e46}) express the lapse and shift functions, 
which are an inherent part of the 4-dimensional metric tensor
 \cite{mtw}, in terms of the
gravitational cylindrical creation and  destruction operators $A(k), \ \ A^+(k)$
which  determine the constituents of the related  GW  and, therefore, the 
geometry it imposes upon the neighbouring space-time. 

\markright{SPACE ZENO EFFECT FOR THE CYLINDRICAL GW}

\protect \section{Space Zeno effect for the cylindrical GW}

For demonstrating the Zeno effect in the cylindrical GW we use
 the Dirac quantization of the cylindrical GW in the half
parametrized formalism as represented in \cite{kuchar1}. We may use the 
 ADM
quantization \cite{kuchar1} but we prefer to discuss the mentioned Dirac
one 
which is more general \cite{kuchar1} than that of ADM. This quantization 
 in the half
parametrized formalism  is equivalent, as shown in \cite{kuchar1} (see Section
XII there),   to the the full parametrized one and so one may discuss the former
formalism without losing anything. In this formalism the function $r(R)$ is
equated to $R$ so that the two remaining canonical coordinates $T(r(R))$, 
$\psi(r(R))$ may be written as $T(r(R))=T(R)$ and $\psi(r(R))=\psi(R)$. These
variables, as well as their canonically conjugate momenta  $\Pi_T(R)$ and
$\pi_{\psi}(R)$, are replaced in the quantum theory \cite{schiff,merzbacher} 
by operators 
so that one may write, for example, the conjugate momenta as the variational
derivatives 

\begin{equation} \Pi_T(R)=-i\frac{\delta}{\delta T(R)}, \ \ \ 
\pi_{\psi}(R)=-i\frac{\delta}{\delta \psi(R)}  \label{e47}
\end{equation}

In the representation in which  the former canonical coordinates $T(R)$ and
$\psi(R)$ are diagonal 
the state functional $\Psi$ depends on these two functions and its behaviour
changes according to the following Schroedinger-type equation

\begin{equation} i\frac{\delta \Psi(T(R),\psi(R))}{\delta
T(R)}={\cal H}(T_{,R}(R),\psi(R),\pi_{\psi}(R))\Psi(T(R),\psi(R)), \label{e48}
\end{equation} 
where in the half parametrized formalism ${\cal H}$ is given by \cite{kuchar1}

\begin{eqnarray}  && {\cal H}=\frac{1}{2}\bigl(1-T^2_{,R}(R)\bigr)^{-1}
\biggl(-iR^{-\frac{1}{2}}\frac{\delta}{\delta \psi(R)}-
R^{\frac{1}{2}}T_{,R}(R)\psi_{,R}(R)\biggr)^2+
\frac{1}{2}R\psi^2_{,R}(R)=   \nonumber \\ && = 
\frac{1}{2(1-T^2_{,R}(R))}\biggl(R^{-1}\pi^2_{\psi}(R)
-2T_{,R}(R)\pi_{\psi}(R)\psi_{,R}(R) +\frac{1}{2}R\psi^2_{,R}(R)\biggr) 
 \label{e49} 
\end{eqnarray}
The  last result was obtained by using the second equation of (\ref{e47}) 
for
the operator $\pi_{\psi}(R)$ and noting that the  commutation relation between 
$\pi_{\psi}(R)$ and $\psi_{,R}(R)$ is zero at the same point, i.g., 
$[\psi_{,R}(R),\pi_{\psi}(R')]=i\frac{\delta(\psi_{,R}(R))}{\delta \psi(R')}=
i\frac{d}{dR}(\frac{\delta(\psi(R))}{\delta \psi(R')})=
i\frac{d\delta(R-R')}{dR}=0$ because of the antisymmetry of the $\delta$
 function by which one have $\frac{d\delta(0)}{dR}=0$ (see the beginning of Section XI in \cite{kuchar1} and the first two unnumbered equations there). 
Note that the variational derivative at the left of (\ref{e48}) is with respect to the
extrinsic time $T(R)$ and $T_{,R}(R)$,  $\psi_{,R}(R)$ denote derivatives of 
$T(R)$ and
$\psi(R)$ with respect to $R$.   One may also note that although 
Eq (\ref{e48}) is 
a functional differential equation which 
 is a whole set
of equations, one for each value of $R$,    it may be discussed 
 for the purpose of this section,
which is the
demonstration of the space Zeno effect for the cylindrical GW  (as emphasized 
  after the following Eq (\ref{e50})), as if it was a 
partial differential equation. This is, especially, true when assigning to $R$
some specific value. In such case  the solution to the
Schroedinger-type  
equation
(\ref{e48}) 
is 
\begin{equation} \Psi(T(R),\psi(R))= 
\exp\bigl(-i{\cal H}(T_{,R}(R),\psi(R))T(R)\bigr)\Psi(0,\psi(R)),  \label{e50}
\end{equation}
where $\Psi(0,\psi(R))$ is the state functional on the hypersurface $T(R)=0$.
Indeed, differentiating $\Psi(T(R),\psi(R))$ from the last equation with respect to
$T(R)$ and taking into account that ${\cal H}(T_{,R}(R),\psi(R))$ does not
depend explicitly on $T(R)$ one obtains the Schroedinger-type equation
(\ref{e48}).  
As emphasized in \cite{kuchar1},    the  set
of equations  for all values of $R$ which are implied by the 
functional differential  Eq (\ref{e48})  are not mutually
independent and may be reduced, as shown in \cite{kuchar1}, to a single
differential equation as follows. 
$$ i\frac{\partial(\bar{\Psi}(T(R,t),\psi(R)))}{\partial
t}=\bar{ H}_t(\psi(R),\pi_{\psi}(R))\bar{\Psi}(T(R,t),\psi(R)), $$ 
where $$\bar{ H}_t(\psi(R),\pi_{\psi}(R))=\int_0^{\infty}dR
{\cal H}(T_{,R}(R,t),\psi(R),\pi_{\psi}(R))\frac{\partial(T(R,t))}{\partial t}$$
The $t$ is a labelling parameter from some closed  range, for example, 
 $[0,1]$ and \\ ${\cal H}(T_{,R}(R,t),\psi(R),\pi_{\psi}(R))$ 
is the same as that of Eq (\ref{e48})-(\ref{e50}). The solution to the former
single differential equation is \cite{kuchar1}  $$  \bar{\Psi}(T(R),\psi(R))= 
P\exp(-i\int_0^1dt\bar{ H}_t)\bar{\Psi}_0(\psi(R)),
$$
where $P$ is a time-ordering operator. 
As mentioned, we prefer,  for 
demonstrating   the space Zeno effect for the cylindrical GW,  to
directly discuss the Hamiltonian density \\
${\cal H}(T(R),\psi(R),\pi_{\psi}(R))$  
and Eq (\ref{e48})   with the understanding
that all the results obtained may be applied also to the  single differential
equation.  This is realized  when noting from the former equations that  $ \bar{\Psi}(T(R),\psi(R))$  is in effect a very large number,  over all values of $R$ and $t$ (in the range $0 \le t \le 1$),  of multiplying products  of the kind of ${\Psi}(T(R),\psi(R))$ from Eq (\ref{e50}).  Thus, since it has been proved in Appendix $B$ and the following equations that for  the function $ {\Psi}(T(R),\psi(R))$ the relevant Zeno  probability  $Pr^{(1)}(|\Psi(0,\psi(R))\!>,\rho)$ tends to unity so it may be seen that for the function  $ \bar{\Psi}(T(R),\psi(R))$ the relevant Zeno probability $Pr^{(1)}(|\bar{\Psi}(0,\psi(R))\!>,\rho)$ becomes  a very large number of multiplying products each tending to the unity value and, therefore, this probability tends as well to this value. This, of course, takes into account that $ \bar{\Psi}(T(R),\psi(R))$ shares the same characteristics with $ {\Psi}(T(R),\psi(R))$, especially, the normalizatiom property (see the discussions before Eq (\ref{$B_{14}$}) and after Eq (\ref{$B_{12}$}) in Appendix $B$). \par
We show here that under the influence of space Zeno effect the state
$\Psi(T(R),\psi(R))$ remain stable in space. 
For that purpose we have calculated in Appendix  $B$,  
using   the shift operator $e^{iP\rho}$,  
the probability $Pr^{(1)}(|\Psi(0,\psi(R))\!>,\rho)$ 
 for the state
 function $\Psi(0,\psi(R))$ from Eq (\ref{e50}) (we also use  the Dirac's bra
 notation \cite{schiff,merzbacher} $|\Psi(0,\psi(R))\!>$)  
 to remain at the same state 
  after shifting  it by the 
 small amount $\rho$. This probability, after the single shift, is  given in
 (\ref{$B_{15}$}) in Appendix $B$ as  
 \begin{align} &   Pr^{(1)}(|\Psi(0,\psi(R))\!>,\rho)=
 \biggl|<\!\Psi(0,\psi(R))|e^{iP\rho}|\Psi(0,\psi(R))\!>\biggr|^2=
 1+4\eta^2+ \nonumber \\ & + 6\eta^4+  
\rho^2\eta^4(\frac{B\pi_{\psi}(R)}{D})^2+
 \eta^2\frac{\rho^2}{D^2}\biggl(2A\pi_{\psi}(R)+B\biggr)^2+ 
 \sin(\frac{\rho}{D}B\pi_{\psi}(R)) \nonumber \\ & \cdot 
 \biggl\{2\eta^3\frac{\rho}{D\pi^2_{\psi}(R)} 
 \biggl(2A\pi_{\psi}(R) + B\biggr)
 -2\eta^2\frac{\rho B\pi_{\psi}(R)}{D}\bigl(1+2\eta^2\bigr)\biggr\}- 
 \label{e51} \\ & -
 \cos(\frac{\rho}{D}B\pi_{\psi}(R))  
 \biggl\{2\eta^3\frac{\rho^2B}{D^2\pi_{\psi}(R)}\biggl(2A\pi_{\psi}(R)+B\biggr)
 +  2\eta^2\bigl(1+2\eta^2\bigr)\biggr\}+ \nonumber \\ & +  
2\eta^3\frac{\rho}{D\pi^2_{\psi}(R)}
\biggl(2A\pi_{\psi}(R)+ B\biggr) \sin(\frac{\rho}{D}A\pi^2_{\psi}(R))
-2\eta^2\bigl(1+2\eta^2\bigr) \nonumber \\ & \cdot  
\cos(\frac{\rho}{D}A\pi^2_{\psi}(R))- 
2\eta^4\frac{\rho B\pi_{\psi}(R)}{D} 
\sin\biggl(\frac{\rho}{D}\biggl(A\pi^2_{\psi}(R)- B\pi_{\psi}(R)\biggr)\biggr)
+ \nonumber \\ & + 2\eta^4\cos\biggl(\frac{\rho}{D}\biggl(A\pi^2_{\psi}(R)-
B\pi_{\psi}(R)\biggr)\biggr)
\nonumber \end{align}
where $A$, $B$, $C$,  $D$ are given by Eqs (\ref{$B_6$}) in Appendix $B$. 
The relevant state  which we want to keep fixed   
 represents the cylindrical GW. 
We, now,  wish to generalize this single shift in one region of space to an
arbitrary number of identical shiftings simultaneously performed in a large
number of independent regions of space.  For that we   construct an operator, 
as done in \cite{bar1}, which
regulate the simultaneous identical shiftings in separete regions of spacetime in
terms of the known quantum projection operator \cite{schiff,merzbacher} 
${\cal
P}=|\phi\!><\!\phi|$ and its translate in shifts of amount $\rho$ which is
denoted here for the single shift as  
$ {\cal P}^{(1)}_{tr}=
e^{-iP\rho}{\cal P}e^{iP\rho}=e^{-iP\rho}<\!\phi|e^{iP\rho}|\phi\!> $. 
 Taking the absolute square of the last expression one obtains 
\begin{equation} |{\cal P}^{(1)}_{tr}|^2=
|e^{-iP\rho}{\cal P}e^{iP\rho}|^2=|<\!\phi|e^{iP\rho}|\phi\!>|^2, \label{e52} 
\end{equation} 
which, assuming for $\phi\!\!>$ the state functional $|\Psi(0,\psi(R))\!\!>$, 
amount
to the probability $Pr^{(1)}(|\Psi(0,\psi(R))\!>,\rho)$ from Eq (\ref{e51}). That is, 
$Pr^{(1)}(|\Psi(0,\psi(R))\!>,\rho)= |{\cal P}^{(1)}_{tr}|^2$. For 
 the double
identical shifts of the state functional $|\Psi(0,\psi(R))\!>$, each of 
amount $\rho$, 
 simultaneously performed in two separate regions of space-time
one may write for ${\cal P}^{(2)}_{tr}$ and its absolute square 
$|{\cal P}^{(2)}_{tr}|^2$
\begin{align} & {\cal P}^{(2)}_{tr}=
e^{-i2P\rho}{\cal P}e^{i2P\rho}
e^{-iP\rho}{\cal P}e^{iP\rho}=
e^{-i2P\rho}{\cal P}e^{iP\rho}{\cal P}
e^{iP\rho}  \label{e53}  \\ &
|{\cal P}^{(2)}_{tr}|^2=\biggl|e^{-i2P\rho}{\cal P}e^{iP\rho}{\cal P}
e^{iP\rho}\biggr|^2=\biggl|e^{-i2P\rho}
\biggl(<\!\Psi(0,\psi(R))|e^{iP\rho}|\Psi(0,\psi(R))\!>\biggr)^2\biggr|^2=  \nonumber \\ 
& =
\biggl|\biggl(<\!\Psi(0,\psi(R))|e^{iP\rho}|\Psi(0,\psi(R))\!>\biggr)^2\biggr|^2,  \nonumber 
\end{align} 
which is identical to the 
probability to remain at the same state  $|\Psi(0,\psi(R))\!>$ 
after these two simultaneous shiftings
performed in two separate regions of space-time. That is, 
$|{\cal P}^{(2)}_{tr}|^2=Pr^{(2)}(|\Psi(0,\psi(R))\!>,\rho)=
\biggl( Pr^{(1)}(|\Psi(0,\psi(R))\!>,\rho)\biggr)^2$.  The generalization to $n$
arbitrary shiftings of $|\Psi(0,\psi(R))\!>$, each of amount $\rho$, 
 in $n$ separate subregions of space,  all included in the finite larger region
 denoted by $X$,
 is
\begin{align} & {\cal P}^{(n)}_{tr}=
e^{-inP\rho}{\cal P}e^{inP\rho}
e^{-i(n-1)P\rho}{\cal P}e^{i(n-1)P\rho}\ldots e^{-iP\rho}{\cal P}
e^{iP\rho} = \nonumber \\ & =
e^{-inP\rho}|\Psi(0,\psi(R))\!>  
\biggl(<\!\Psi(0,\psi(R))|e^{iP\rho}|\Psi(0,\psi(R))\!>\biggr)^{n-1} 
\label{e54} \\ & 
\cdot <\!\Psi(0,\psi(R))|e^{iP\rho} 
=e^{-inP\rho}\biggl(<\!\Psi(0,\psi(R))|e^{iP\rho}|\Psi(0,\psi(R))\!>\biggr)^n
  \nonumber 
\end{align} 
The absolute square of the last expression is equal to the probability to remain
at the same state after $n$ simultaneous shiftings, each of amount $\rho$, 
in $n$ separate independent regions of spacetime. That is, raising the 
probability 
$Pr^{(1)}(|\Psi(0,\psi(R))\!>,\rho)$ from Eq (\ref{e51}) to the $n$-th power one obtains 
\begin{align} & |{\cal P}^{(n)}_{tr}|^2=
\biggl|e^{-inP\rho}|\Psi(0,\psi(R))\!>\biggl(<\!\Psi(0,\psi(R))|e^{iP\rho}|\Psi(0,\psi(R))\!>)^{n-1}
<\!\Psi(0,\psi(R))|e^{iP\rho}\biggr)\biggr|^2 =\nonumber \\  & 
=\biggl|\biggl(<\!\Psi(0,\psi(R))|e^{iP\rho}|\Psi(0,\psi(R))\!>\biggr)^n\biggr|^2=
\biggl(Pr^{(1)}(|\Psi(0,\psi(R))\!>,\rho)\biggr)^n= \nonumber \\ & = 
\biggl\{  1+4\eta^2+
 6\eta^4+  
\rho^2\eta^4(\frac{B\pi_{\psi}(R)}{D})^2+
 \eta^2\frac{\rho^2}{D^2}\biggl(2A\pi_{\psi}(R)+B\biggr)^2+\label{e55} \\ & +
 \sin(\frac{\rho}{D}B\pi_{\psi}(R))  
 \biggl\{2\eta^3\frac{\rho}{D\pi^2_{\psi}(R)}\biggl(2A\pi_{\psi}(R) + B\biggr)
 -2\eta^2\frac{\rho B\pi_{\psi}(R)}{D}\bigl(1+2\eta^2\bigr)\biggr\}- \nonumber
 \\ & - 
 \cos(\frac{\rho}{D}B\pi_{\psi}(R))
 \biggl\{2\eta^3\frac{\rho^2B}{D^2\pi_{\psi}(R)}\biggl(2A\pi_{\psi}(R)+B\biggr)
 + 2\eta^2\bigl(1+2\eta^2\bigr)\biggr\}+ \nonumber \\ & +  
2\eta^3\frac{\rho}{D\pi^2_{\psi}(R)}
\biggl(2A\pi_{\psi}(R)+B\biggr)\sin(\frac{\rho}{D}A\pi^2_{\psi}(R))
 - 2\eta^2\bigl(1+2\eta^2\bigr)
\cos(\frac{\rho}{D}A\pi^2_{\psi}(R))- \nonumber \\ & - 
2\eta^4\frac{\rho B\pi_{\psi}(R)}{D}
\sin\biggl(\frac{\rho}{D}\biggl(A\pi^2_{\psi}(R)-B\pi_{\psi}(R)\biggr)\biggr)
+  2\eta^4\cos\biggl(\frac{\rho}{D}\biggl(A\pi^2_{\psi}(R)-
B\pi_{\psi}(R)\biggr)\biggr) \biggr\}^{\frac{X}{\rho}}, 
\nonumber \end{align} 
where  the $n$, which refers to the $n$ equal shiftings each of amount
$\rho$, 
  were written as $\frac{X}{\rho}$ 
in which $X$ is the finite region of
space which includes all the $n$ smaller subregions. 
 We, now, approach the Zeno limit   of
$\lim_{\rho \to 0}$ and note that in this limit one have  
\begin{align} & \lim_{\rho \to 0}\biggl( \cos(\frac{\rho}{D}
A\pi^2_{\psi}(R))\biggr)=\lim_{\rho \to 0} \biggl(\cos(\frac{\rho}{D}
B\pi_{\psi}(R))\biggr)= \nonumber \\ & = 
\lim_{\rho \to 0}\biggl(\cos\biggl(\frac{\rho}{D}\biggl(A\pi^2_{\psi}(R)-
B\pi_{\psi}(R)\biggr)\biggr)\biggr)=1 \nonumber \\  & 
\lim_{\rho \to 0}\biggl( \sin(\frac{\rho}{D}
A\pi^2_{\psi}(R))\biggr)=\lim_{\rho \to 0} \biggl(\sin(\frac{\rho}{D}
B\pi_{\psi}(R))\biggr)=  \nonumber \\ & =
\lim_{\rho \to 0}\biggl(\sin\biggl(\frac{\rho}{D}\biggl(A\pi^2_{\psi}(R)-
B\pi_{\psi}(R)\biggr)\biggr)\biggr)=0 \nonumber 
\end{align}
Thus, using the last limits,  one obtains for 
 the overall probability $(Pr^{(1)}(|\Psi(0,\psi(R))\!>,\rho))^n$  from Eq
 (\ref{e55})  in the
limit $\rho \to 0$
\begin{equation} \label{e56} \lim_{\rho \to 0}
\biggl(Pr^{(1)}(|\Psi(0,\psi(R))\!>,\rho)\biggr)^{\frac{X}{\rho}}  =1, 
 \end{equation}
where we use the mentioned equality $n=\frac{X}{\rho}$. Note that the last
result is obtained for any value of the eigenvalue $\eta$. Thus, it is shown
that in the limit of space Zeno effect the cylindrical GW,  which is
interpreted in the framework of quantum mechanics in probability terms, have a
unity probability for being found in some spacetime region and, therefore,  if it
is strong enough for implanting its geometry upon this neighbourhoud.  

\markright{GRAPHICAL REPRESENTATION OF THE GRAVITATIONAL SPACE ZENO....}
 
\protect \section{ Graphical representation of the gravitational 
 space Zeno effect}
 
We, now, wish to corroborate our former results through numerical and graphical
representation. For that we should note that the idea of the space Zeno 
effect,  which is to
remain with the same cylindrical geometry across all space,  necessitates a
constancy of the function $T(R)$ across all the $n$ subregions of space. 
Thus,  
one may assume $\frac{dT(R)}{dR} \approx 0$ which entails 
$\frac{d^2T(R)}{dR^2} = (\frac{dT(R)}{dR})^2 =0$. In such case the functions
$A$, $B$, $C$, $E$ from Eqs (\ref{$B_6$}) assume a very simple form. That is 
\begin{eqnarray} && A_{(T_{,R}=0)}=T(R), \ \ \ \ \ \ \ \ \ \ \  B_{(T_{,R}=0)}=0  
\label{e57} \\ && 
 C_{(T_{,R}=0)}=- T(R)
\biggl(\frac{R^3}{2}+  R\rho(R+\frac{\rho}{2})\biggr)
\biggl(\rho(\frac{d^2\psi(R)}{dR^2})^2  
+2\psi_{,R}(R)\frac{d^2\psi(R)}{dR^2}\biggr) - \nonumber \\ && 
-\frac{R}{2}(R+\rho)T(R)\psi^2_{,R}(R) \nonumber \\ && 
D_{(T_{,R}=0)}=2R(R+\rho) \nonumber 
\end{eqnarray} 
Substituting from the last $A_{(T_{,R}=0)}$, $B_{(T_{,R}=0)}$ and 
$D_{(T_{,R}=0)}$ in Eq (\ref{e55})  for the probability one obtains 
\begin{eqnarray} &&  
\biggl(Pr^{(1)}_{(T_{,R}=0)}(|\Psi(0,\psi(R))\!>,\rho)\biggr)^{\frac{X}{\rho}}=
\biggl(\biggl|<\!\Psi(0,\psi(R))|e^{iP\rho}
|\Psi(0,\psi(R))\!>\biggr|_{(T_{,R}=0)}^2\biggr)^{\frac{X}{\rho}}= \nonumber \\ && = 
 \biggl(1+4\eta^2+  6\eta^4+  
 4\eta^2\frac{\rho^2}{D^2}\pi^2_{\psi}(R) 
- 2\eta^2\bigl(1+2\eta^2\bigr)- \label{e58} \\ && -2\eta^2(1+\eta^2)
\cos(\frac{\rho}{D}T(R)\pi^2_{\psi}(R))  + 
4\eta^3\frac{\rho}{D\pi_{\psi}(R)}
\sin(\frac{\rho}{D}T(R)\pi^2_{\psi}(R))\biggr)^{\frac{X}{\rho}}
\nonumber \end{eqnarray}
For $\psi(R)$ we use the quantum version of Eq (\ref{e8}) for  
a specific value of $k$ and
for $\pi_{\psi}(R)$ we 
 take  into account that we discuss here $T_{,R}=0$ so one may write 
 the quantum
 version of 
 $\pi_{\psi}(R)$ from Eq (\ref{e13}) as 
 \begin{equation}   \label{e59}  \pi_{\psi}(R)=iR k
J_0(kR)\bigl(A(k)e^{(ikT)}-A^+(k)e^{-(ikT)}\bigr)R_{,r},   
\end{equation} 
where $R_{,r}=1$ because $R=r$ here. In 
the equality $n=\frac{X}{\rho}$,   which relates $\rho$ to the
number of subregions $n$ included in the larger region $X$,  we assume 
 $X=20R$. Note the difference between the amplitudes $A(k)$, $A^+(k)$ and $A$
 given by Eq (\ref{e57}) and by the first of Eqs (\ref{$B_6$}). 
We also assume, for avoiding complex expressions and 
 easing the graphical introduction of our results, that the amplitude
$A(k)$ is equal to its complex conjugate $A^{+}(k)$   so that 
  multiplying  the last equation (\ref{e59}) by its conjugate and taking 
   the square root of the
resulting product one  obtains 
\begin{equation}  \label{e60} 
\pi_{\psi}(R)=RkA(k)J_0(kR)\biggl(2\bigl(1-cos(2kT)\bigr)\biggr)^{\frac{1}{2}}   
\end{equation} 
Substituting the last expression for $\pi_{\psi}(R)$ in Eq (\ref{e58}) for the
probability  
 and graphing the resulting expression for certain values of $T$,
$k$, $A$ and $R$ one may visually see how the probability approach unity as
$\rho$ approach zero. Note that although a probability should be in the range of
$(0, 1)$  we obtain here some graphs which exceed unity because of the
caclculations, such as that for $\pi_{\psi}(R)$ in Eq (\ref{e60}), 
 used to assign values to the parameters required for these graphs.
  One  may also realize,   for all the Figures and  Subfigures shown here,    
that the  drawn graphs and surfaces  are at the neighbourhood of
unity  
for the assumed   ranges of
$\rho$ and all approach unity as $\rho \to 0$.
 We note that although $A(k)$ $A^+(k)$ depend upon $k$ we refer to them,
 for the present numerical discussion, as constants and asuume $A=A^+$. 
 Also,  since, as mentioned,  $A$ and $A^+$  are thought 
  in the quantum regime to be  creation and destruction operators and
since these, respectively,  raise and decrease the relevant energy by no more
than unity the corresponding values assumed for  $A$, $A^+$ in the numerical
discussion here should be in the order of magnitude of unity. \par 
  In Figure 1 we see two Subfigures which show the
probability from Eq (\ref{e58}) as function of $\rho$ in the range $1 \geq \rho
\geq 0$ and for $k=1$, $R=100$, $T=1$
and $a=1$. The left Subfigure $1,a$  is drawn for $\eta=1$ and the right
one   for $\eta=9$.    One may  realize that for small
values of the eigenvalue $\eta$, such as in Subfigure $1,a$,   the probability
 is linear whereas for relatively larger values of $\eta$, such as in
Subfigure $1,b$, it deviates from linearity. \par 
In Figure 2 we
show 
three Subfigures which show the
probability from Eq (\ref{e58}) as function of $\rho$ in the range 
$0.2 \geq \rho
\geq 0$, for $k=\frac{1}{2}$, $T=6$, $\eta=1$, $A=A^+=1$  and 
 for certain different
ranges of $R$. Thus, in Subfigure $(1,a)$ we see this probability for 
$0.2 \geq \rho \geq 0$,  $20 \geq R \geq 10$ and one may
see that as $\rho$ becomes smaller  the probability approach unity for all
values of $R$.  Also, one may realize that the larger becomes $R$, even for the
larger values of the drawn range of $\rho$, the probability tends to the unity
value.  Similar situations are also shown in Subfigures $(2,b)$ and $(2,c)$
which are, respectively, drawn for intermediate $120 \geq R \geq 100$ and 
larger values $1020 \geq R \geq 1000$ of $R$ where 
it is clearly shown that the surfaces drawn for these $R$'s have, essentially, a
unity value for the whole range of $0.2 \geq \rho \geq 0$ and become proper
unity as $\rho \to 0$.  \par    In Subfigure $(3,a)$ of Figure 3  we show a
three-dimensional surface of the 
probability from Eq (\ref{e58}) as function of $\rho$ in the range 
$0.5 \geq \rho \geq 0$,  of the eigenvalue $\eta$ in the range 
$2 \geq \rho \geq 0$ and for $k=\frac{1}{2}$, $T=6$, $R=150$ and $A=A^+=1$.
As realized from Eqs (\ref{e56}) even  the general probability from Eq 
(\ref{e55}) tends to
unity as $\rho \to 0$ for any value of $\eta$. We have, nevertheless, show how
the probability 
 from Eq
(\ref{e58}) approach unity in this limit. As seen in  panel $(3,a)$ 
 for a value of $\rho$  
which is very close to
zero and for certain values of $\eta$ the probability discontinuously 
 decreases to  zero and immediately as $\rho$ becomes zero it 
 increases, for the same values of $\eta$, to unity.  Subfigure  $(3,b)$ shows
again this behaviour of  the probability  for the comparatively  extended 
ranges of $1 
\geq \rho \geq 0$,  $9 
\geq \eta \geq
0$ and for $R=100$ whereas $k$, $T$ and $A=A^+$ remain as for Panel $(3,a)$. 
One may again see how for a $\rho$ just before zero the probability
discontinuously fall towards zero and then at $\rho=0$ this probability
discontinuously increases to unity. In Subfigure $(3,c)$ we see
 the probability from Eq (\ref{e58}) as function of $\rho$ and $T$ in the
 respective ranges of  
 $1 \geq \rho \geq 0$ and $100 \geq T \geq 0$ and for 
 $k=\frac{1}{2}$,  $R=100$, $\eta=1$ and $A=A^+=1$.

\begin{figure}
a
\begin{turn}{-90}
\includegraphics[width=5.6cm]{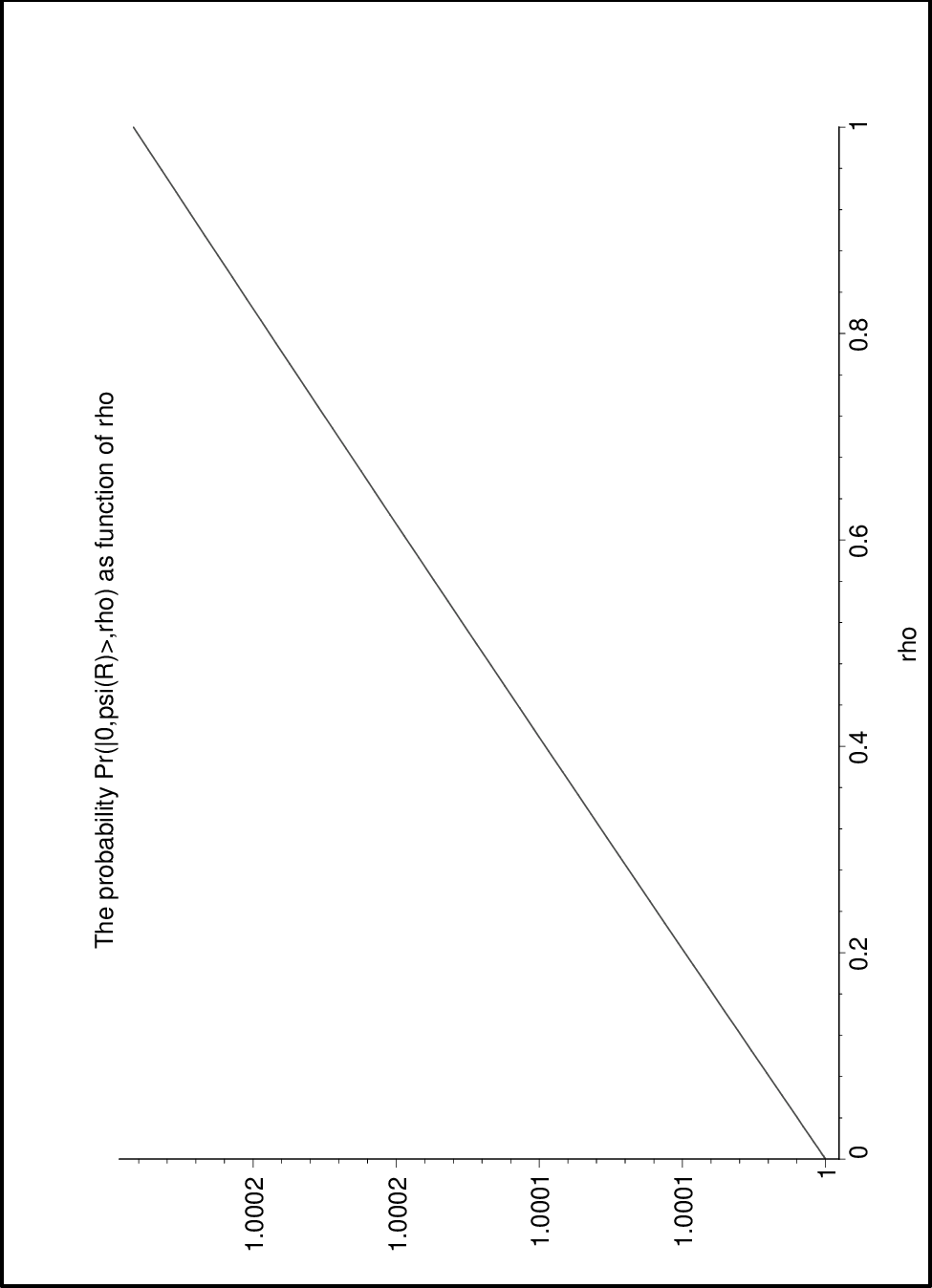}
\end{turn}
b
\begin{turn}{-90}
\includegraphics[width=5.6cm]{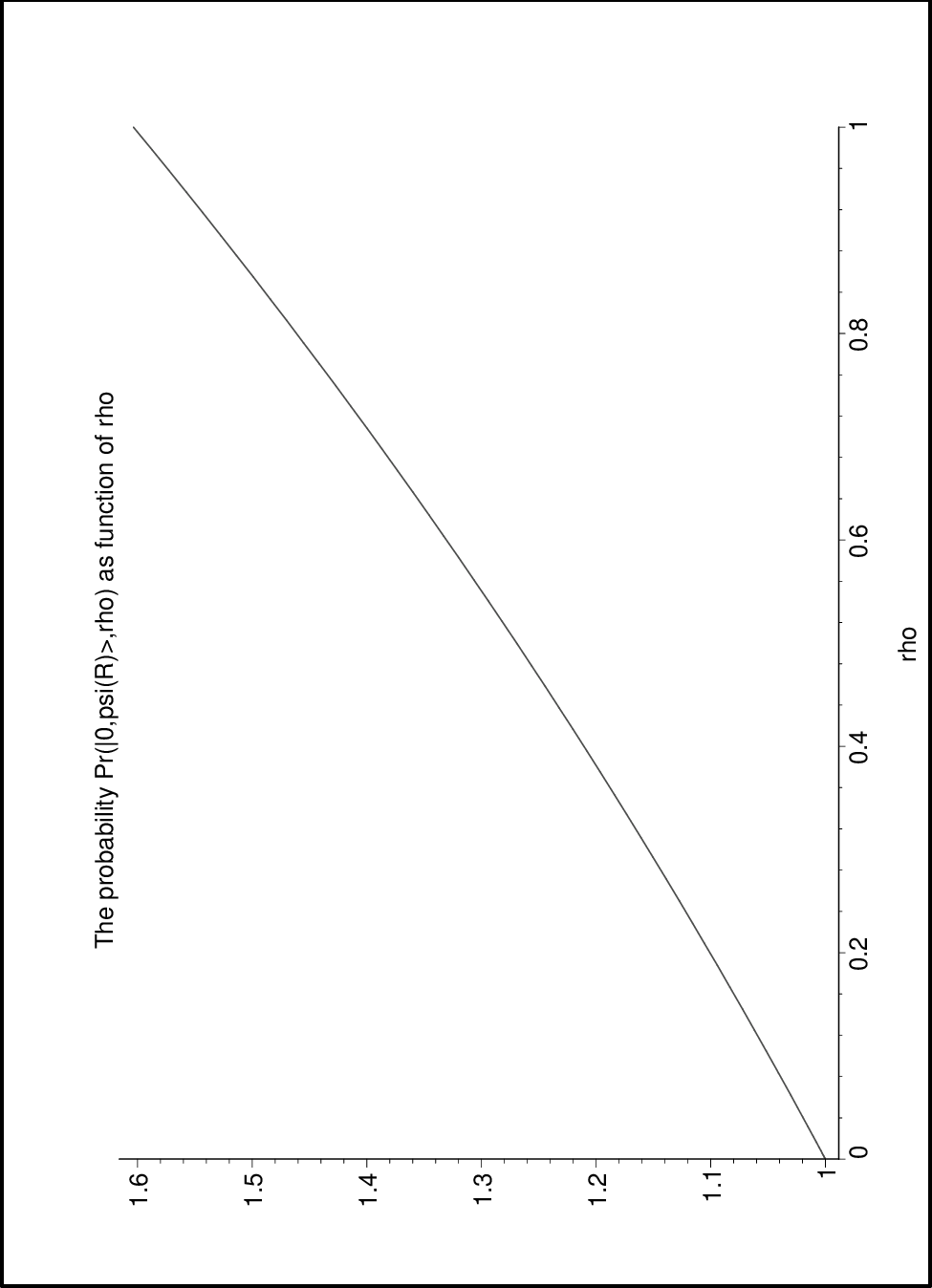}
\end{turn}

\caption{The left panel $a$ shows the probability from Eq (\ref{e58}) as function
of $\rho$ in the range $1 \geq \rho \geq 0$ and for  
$k=1$,  $R=100$, $\eta=1$, $T=1$  and $A=A^+=1$. The right panel $b$ shows this
probability for the same values of the former parameters except for $\eta=9$.
One may see how the graph changes its linear form,  which is valid for 
small $\eta$, to
the nonlinear one as $\eta$ relatively increases.  See also the discussion after 
Eq (\ref{e60}). }

\end{figure}

\begin{figure}
a
\begin{turn}{-90}
\includegraphics[width=5.6cm]{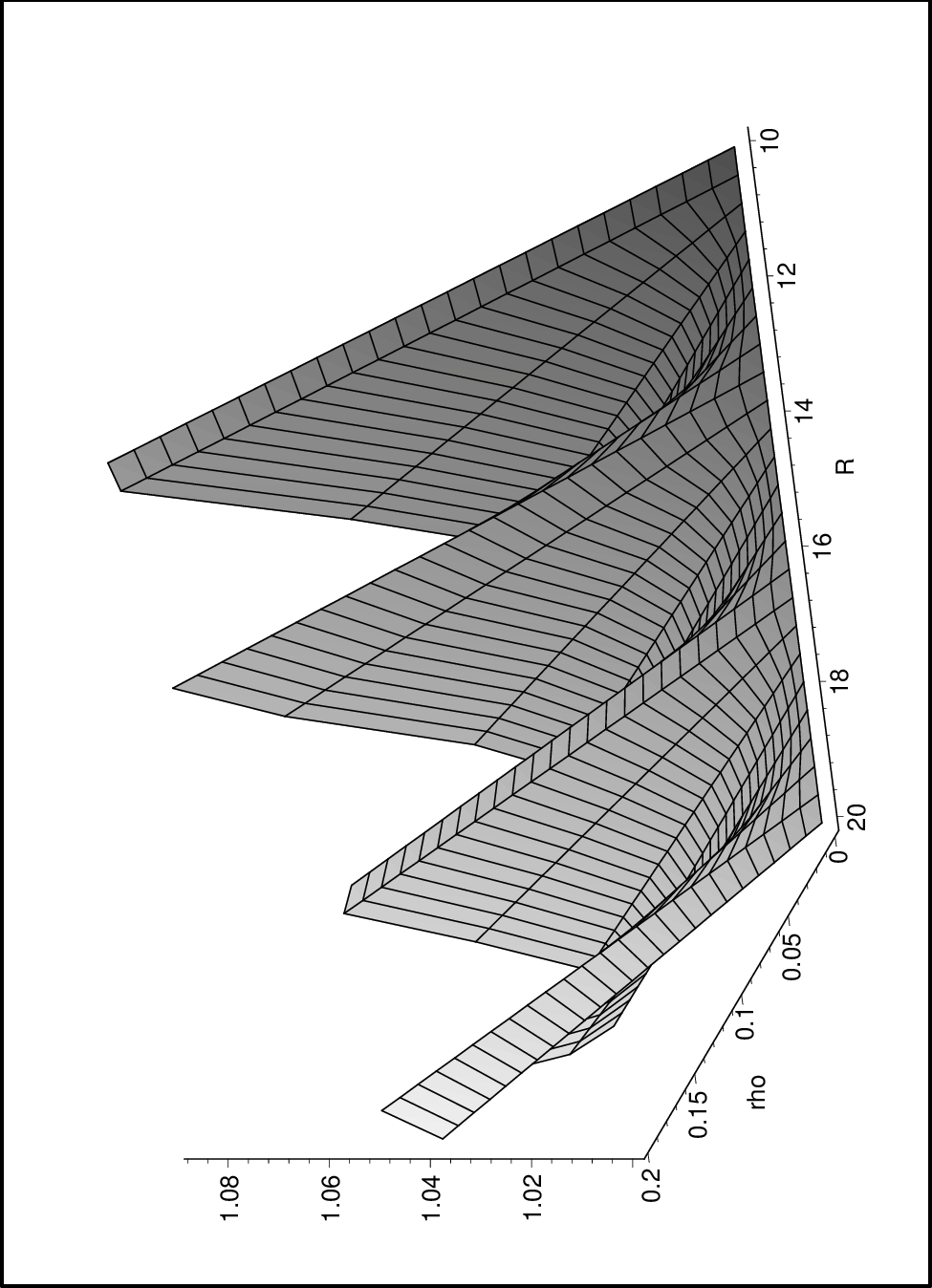}
\end{turn}
b
\begin{turn}{-90}
\includegraphics[width=5.6cm]{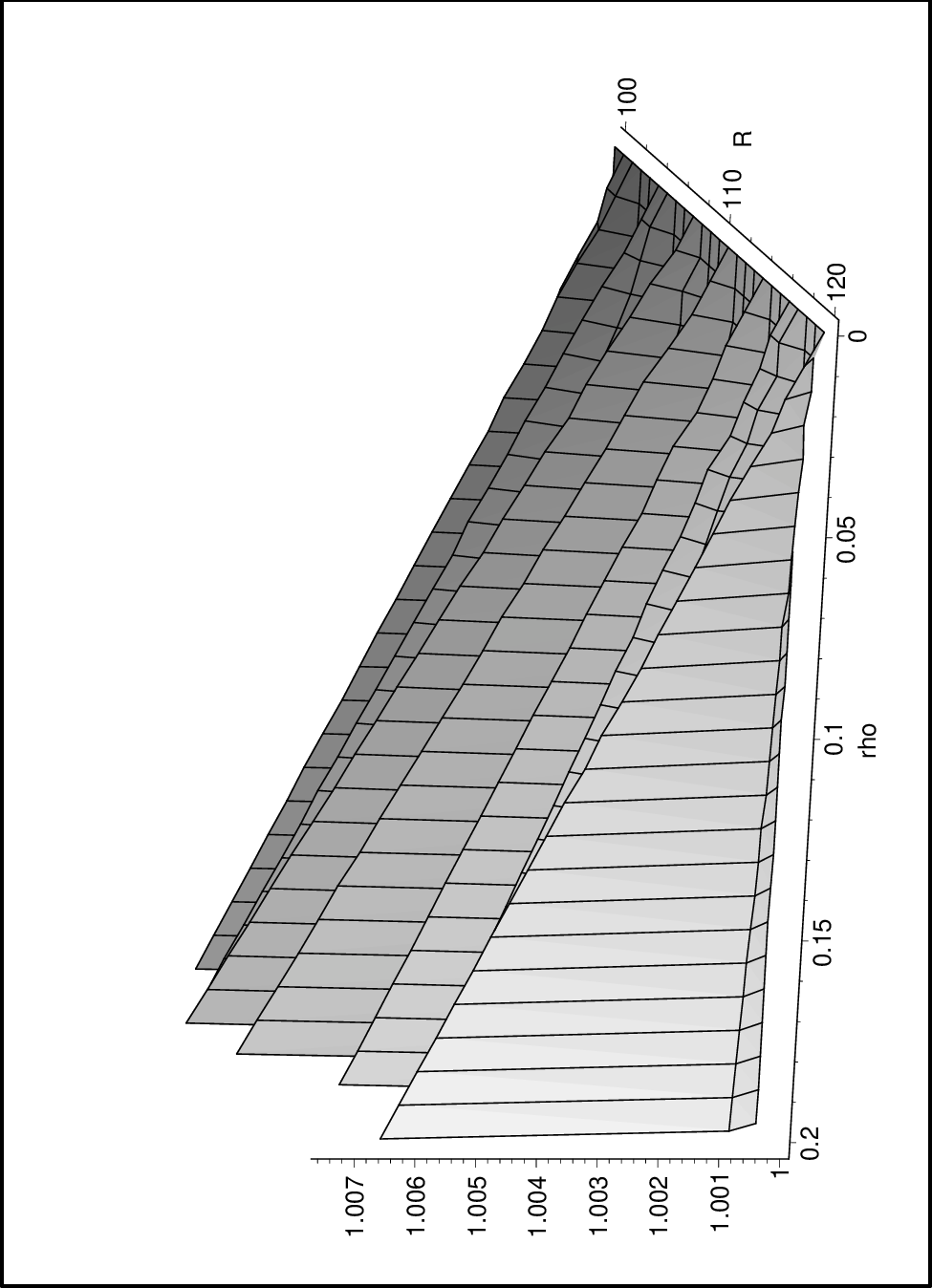}
\end{turn}
c
\begin{turn}{-90}
\includegraphics[width=5.6cm]{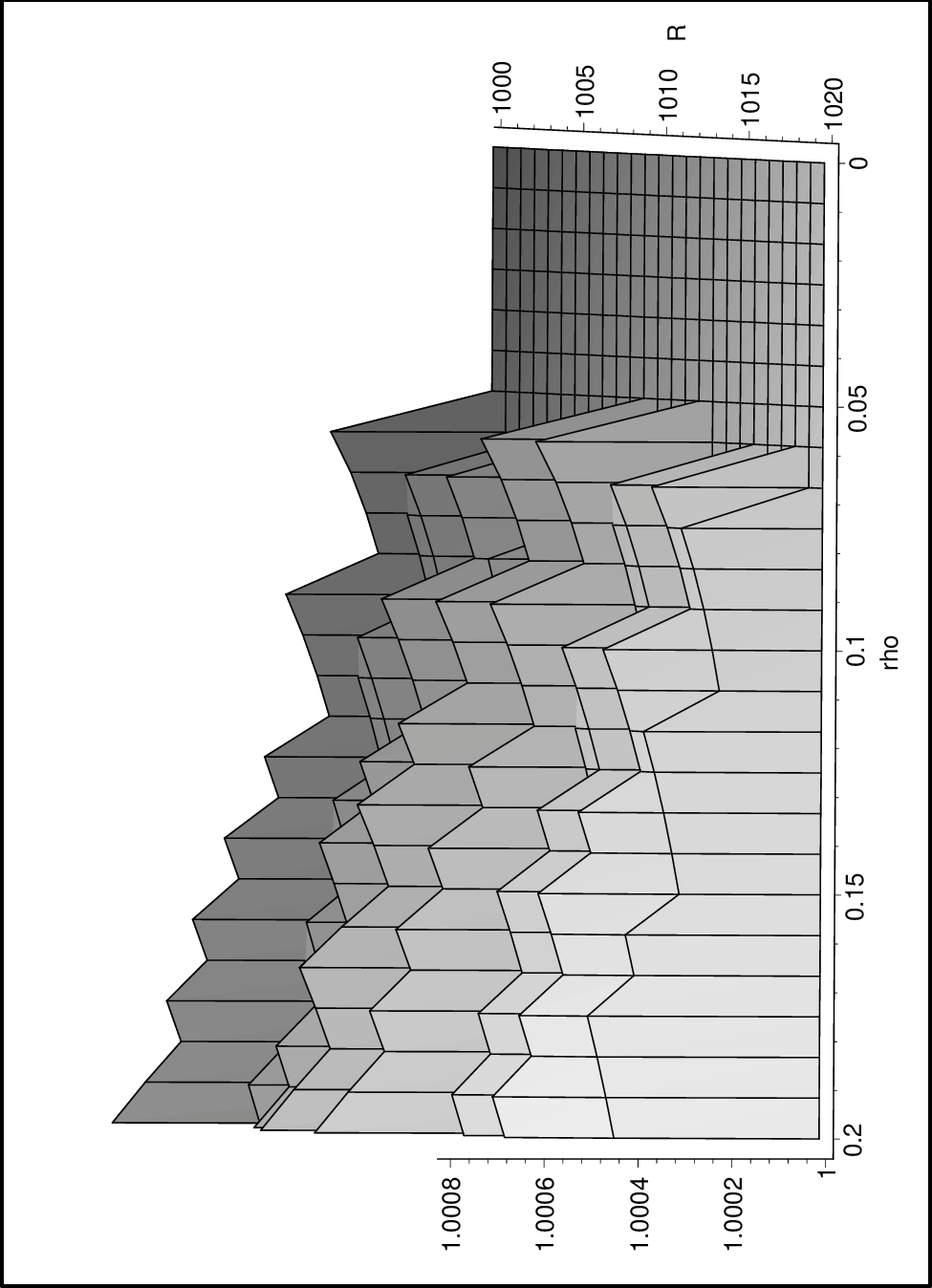}
\end{turn}

\caption{The three panels of this figure show three-dimensional surfaces of the
probability from Eq (\ref{e58}) as function of $\rho$ in the range 
$0.2 \geq \rho \geq 0$, for $k=\frac{1}{2}$, $T=6$, $\eta=1$, $A(k)=A^+=1$ and
for 3 different ranges of $R$. As seen,  for all the three panels the relevant
probability tends to unity as $\rho \to 0$. In Panel $a$ the range of $R$ is 
$20 \geq R \geq 10$ and those of Panels $b$ and $c$ are respectively 
$120 \geq R \geq 100$  and $1020 \geq R \geq 1000$. 
Note that the surfaces shown
in these panels  and also in Subfigure $3,c$ have,   essentially, a unity
value for all their ranges.  }

\end{figure}

\begin{figure}
a
\begin{turn}{-90}
\includegraphics[width=5.6cm]{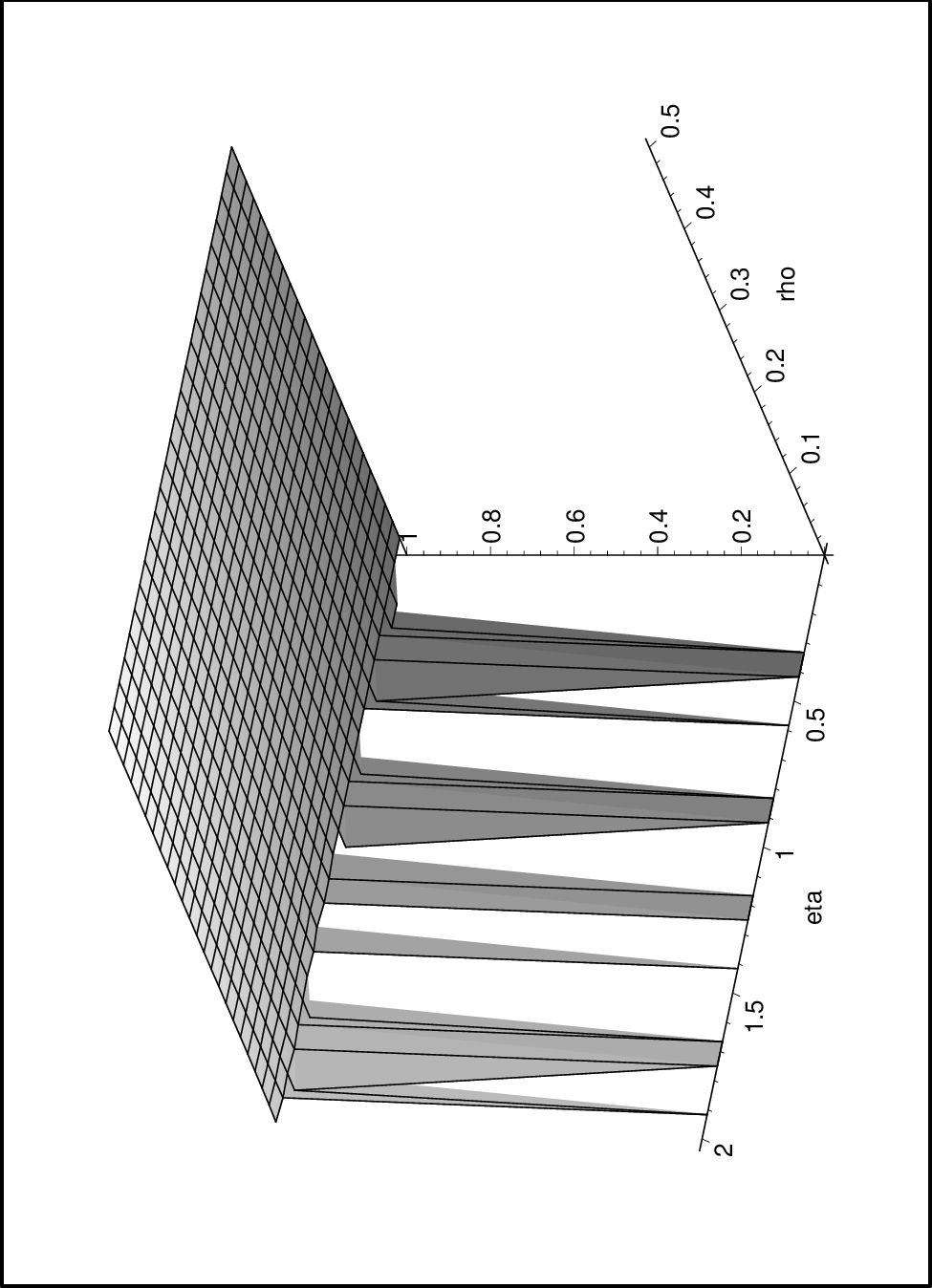}
\end{turn}
b
\begin{turn}{-90}
\includegraphics[width=5.6cm]{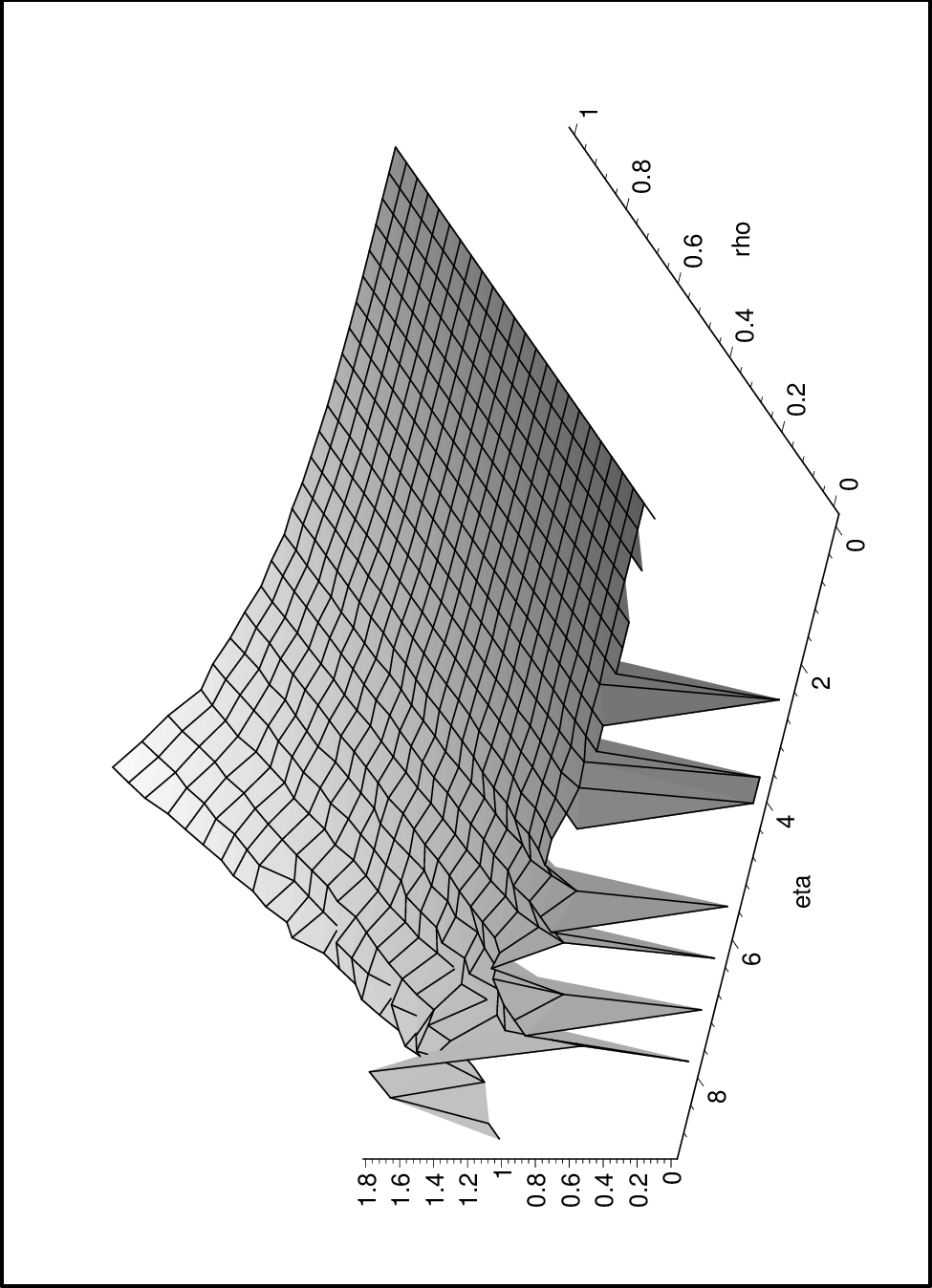}
\end{turn}
c
\begin{turn}{-90}
\includegraphics[width=5.6cm]{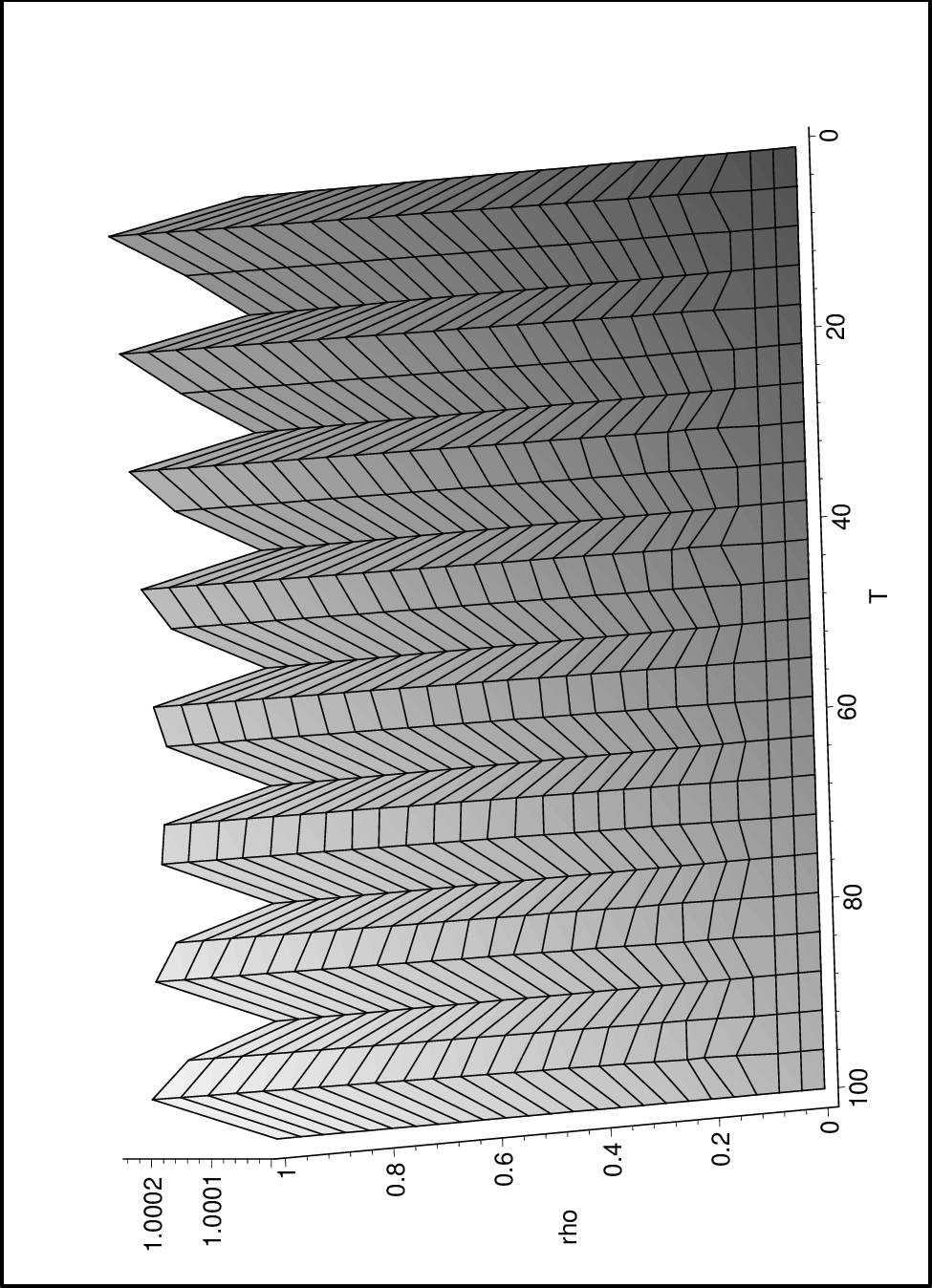}
\end{turn}
\caption{The  panel $a$ shows a three-dimensional surface of 
the probability  from Eq (\ref{e58}) as
function of $\rho$ in the range $0.5 \geq \rho \geq 0$,  $\eta$ in the range 
$2 \geq \rho \geq 0$ and for $k=\frac{1}{2}$, $T=6$, $R=150$ and $A=A^+=1$ . Note that for a value of $\rho$ which is very close to
zero and for certain values of $\eta$ the probability is practically zero and
discontinuously increases to unity as $\rho$ becomes zero.  Panel $b$ shows
again this probability but now for the relatively extended ranges of $1 
\geq \rho \geq 0$,   $9 
\geq \eta \geq
0$ and for $R=100$ whereas $K$, $T$ and $A$, $A^+$ remain as for Panel $a$. One
may again see the discontinuous jumps in the probability just before and at 
$\rho=0$.  
Panel $c$ 
shows a three-dimensional surface of the 
 probability  from Eq (\ref{e58}) as
function of $\rho$ in the range $0.5 \geq \rho \geq 0$ and $T$ in the range 
$100 \geq \rho \geq 1$. One may discern the cyclicity and monotony by 
 which the probability changes with $T$ in the neighbourhood of unity}. 
\end{figure}

\markright{THE CYLINDRICAL GRAVITATIONAL WAVE  IN THE TT GAUGE}

\protect \section{The cylindrical GW in the TT gauge}

We discuss here a linearized version of general relativity in which the
cylindrical GW is considered as a small pertubation in the otherwise flat
Minkowskian metric. That is, the appropriate metric tensor  may be written as 
\begin{equation} g_{\mu\nu}= \label{e61} \epsilon_{\mu\nu}+h_{\mu\nu}, \ \ \ 
|h|<<1, \end{equation} 
where the GW is identified with the small perturbation $h$. 
Thus, the exponents in the cylindrical metrics of Eq
(\ref{e4}) may be expanded in a Taylor series in which we retain only the first
two terms so that taking into account the flat Minkowskian background (in
cylindrical coordinates \cite{mtw}) 
($\epsilon_{\mu\nu}=diag(-1,1,R^2,1)$)  one 
  may  write the metrics  as 
\begin{equation} \label{e62} ds^2=(1+1+
\Gamma-\psi)(-dT^2+dR^2)+(2+2ln(R^2)-\psi)d\phi^2+(1+1+\psi)dz^2,  
\end{equation} 
where the $d\phi^2$ term was obtained by expanding  the exponentials related to both the cylindrical
perturbation $h_{\phi\phi}$ and the Minkowskian background  $\epsilon_{\phi\phi}$  [use was made of  a Taylor series in which 
the first two terms were retained] as follows
\begin{eqnarray*} && ds^2_{\phi\phi}=\epsilon_{\phi\phi}+h_{\phi\phi}=\biggl(R^2+R^2e^{-\psi}\biggr)d\phi^2=
\biggl(e^{\ln(R^2)}+e^{\bigl(\ln(R^2)-\psi\bigr)}\biggr)d\phi^2=  \nonumber \\ && = \biggl(1+\ln(R^2)+
1+\ln(R^2)-\psi\biggr)d\phi^2=  \biggl(2+2\ln(R^2)-\psi\biggr)d\phi^2
\nonumber \end{eqnarray*}
From Eqs (\ref{e61})-(\ref{e62}) one may determine the components $h_{\mu\nu}$ 
 which characterize the cylindrical GW. 
 \begin{equation} \label{e63}  h_{00}=-(1+\Gamma-\psi), \ \ 
 h_{11}=(1+\Gamma-\psi), 
 \ \ 
 h_{22}=(1+ln(R^2)-\psi), \ \ h_{33}=1+\psi, \end{equation} 
 where $\psi$ and $\Gamma$ are respectively given by Eqs (\ref{e9}) and 
 (\ref{e20})-(\ref{e21})  and the suffixes $(0,\ 1,\ 2,\ 3)$ are $(t, \ r, \ \phi, \ z)$. 
Now, since we discuss here pure  GW's in which the matter terms in Einstein's 
field
equations are zero we may use \cite{mtw} the transverse traceless (TT) gauge
\cite{thorne,mtw} which is characterized with a minimum
number of components \cite{mtw} for 
 the metric tensor $h_{\mu \nu}$ so that   any component of it,  
  except the spatial ones,    vanishes, i.e; $h^{TT}_{\mu 0}=0$. 
   The TT gauge is  also characterized with the following properties \cite{mtw}: 
     $h^{TT}_{kj,j}=0$, so that  these components are
divergence-free and  are also trace-free, e.g.,   
$h^{TT}_{kk}=0$. Thus, since, as mentioned, 
the GW is identified with   
$h_{jk}^{TT}$ it, naturally,  shares the same properties. In the linearized
version of general relativity  the components of the Riemann curvature tensor 
are found to be \cite{mtw} 
\begin{equation} R_{\alpha\mu\beta\nu}= \frac{1}{2}\biggl(
h_{\alpha\nu,\mu\beta}+h_{\mu\beta,\nu\alpha}-h_{\mu\nu,\alpha\beta}-
h_{\alpha\beta,\mu\nu}\biggr) \label{e64} \end{equation} 
Also,  it has been shown (see Eq (35.10) in \cite{mtw}) that in the TT gauge the
time-space components of   the Riemann curvature tensor 
  have an especially
simple form (see Eq (35.10) in \cite{mtw})
\begin{equation} \label{e65}  
R_{j0k0}=R_{0j0k}=-R_{j00k}=-R_{0jk0}=-\frac{1}{2}h^{TT}_{jk,00}
\end{equation}
Thus, substituting in (\ref{e64}) $\alpha=j$, $\beta=k$, and $\mu=\nu=0$ one 
obtains 
\begin{equation} \label{e66}
R_{j0k0}=\frac{1}{2}\biggl(h_{j0,0k}+h_{0k,0j}-h_{00,jk}-h_{jk,00}\biggr) 
\end{equation}
From the last two equations (\ref{e65})-(\ref{e66}) and  the vanishing of 
any nondiagonal element 
of the metric tensor $h$ (see Eq (\ref{e63}))  one may realize that if
$h_{00}=0$ then $h_{jk}=h^{TT}_{jk}$ which means that in the linearized version 
of general relativity the cylindrical GW can
be expressed only in the TT gauge. But since, as seen from Eq (\ref{e63}), 
  $h_{00} \neq 0$ one may use Eqs (\ref{e65})-(\ref{e66}) to find  an explicit 
expression for  $h^{TT}_{jk}$
\begin{equation}  \label{e67}
h^{TT}_{jk}=c_1+c_2t+h_{jk}+\int\!dt\int\!dt'h_{00,jk},
\end{equation}
where, as realized from Eq (\ref{e63}),   $j=k$ and $c_1$, $c_2$ are constants
of integration.   Thus, substituting in Eq
(\ref{e67}) for $h_{jk}$ and $h_{00}$ from Eq (\ref{e63}), taking into 
account that these $h$'s depend  only upon $t$ and $r$ 
(through $\psi$, the $\Gamma$ does not depend  upon them as seen from  Eqs
(\ref{e20})-(\ref{e21})),  using Eqs (\ref{e9}) and (\ref{e20})  
for $\psi$ and $\Gamma$, performing the double integration over $t$  
 and using the Bessel's derivatives \cite{abramowitz} 
$\frac{dJ_0(x)}{dx}=-J_1(x)$,
$\frac{dJ_1(x)}{dx}=\frac{1}{2}(J_0(x)-J_2(x))$   
 one may obtain for the components of $h^{TT}_{jk}$ 
\begin{align}  &
h^{TT}_{rr}=\Re\biggl\{\biggl(c_1+c_2t+h_{rr}+\int\!dt\int\!dt'h_{00,rr}\biggr)
\biggr\}=
\Re\biggl\{\biggl(c_1+c_2t+\Gamma-\psi+ \nonumber \\ & + 
\int\!dt\int\!dt'(\psi-\Gamma-1)_{,rr}\biggr)\biggr\} =
\Re\biggl\{c_1+c_2t+\Gamma-\frac{1}{2}\int_0^{\infty}
dk\biggl(J_0(kr)+J_2(kr)\biggr) \cdot \nonumber \\ & \cdot 
\biggl(A(k)e^{ikT}+A^+(k)e^{-ikT}\biggr)\biggr\}=
 c_1+c_2t-\frac{1}{2}\biggl\{\int_0^{\infty}dk
 \biggl[\bigl(J_0(kr)+J_2(kr)\bigr) \cdot \label{e68} \\ & \cdot 
 \biggl(A(k)+A^+(k)\biggr)\cos(kT)-
4kA(k)A^+(k)\biggr]\biggr\} \nonumber 
\end{align}
\begin{align}   &  
h^{TT}_{\phi\phi}=\Re\biggl(c_1+c_2t+h_{\phi\phi}\biggr)=\Re\biggl(c_1+c_2t+
\ln(R^2)-\psi\biggr)=
\label{e69} \\ & =
c_1+c_2t+\ln(R^2)-\biggl(\int_0^{\infty}J_0(kr)\cos(kT)(A(k)+A^+(k))\biggr)
\nonumber \end{align}
\begin{equation}  \label{e70}
h^{TT}_{zz}=\Re\biggl(c_1+c_2t+h_{zz}\biggr)=
c_1+c_2t+\int_0^{\infty}J_0(kr)\cos(kT)(A(k)+A^+(k)),
\end{equation} 
where the unity terms of Eqs (\ref{e63}) are included in the constants $c_1$ and
$\Re$ is the real part of the relevant expressions. These components of the cylidrical GW are,
naturally, polarized along their respective directions just as the
electromagnetic waves are polarized along their spatial axes. We should, 
however, 
take into account that these GW's are tensors and therefore their polarization
directions have, likewise, tensorial \cite{synge} characters. The corresponding unit
polarization tensors were introduced in \cite{mtw},  with respect to plane GW
propagating along a general $\hat{n}$ axis  in an orthogonal 
  $\hat{n}, \hat{n_1}, \hat{n_2}$ 
system, as 
\begin{align} & {\bf e}_{+_{\hat{{\bf n}}_1\hat{{\bf n}}_1}}= 
 {\bf e}_{\hat{{\bf n}}_1} \otimes {\bf e}_{\hat{{\bf n}}_1}- {\bf e}_{\hat{{\bf
 n}}_2} \otimes {\bf e}_{\hat{{\bf n}}_2}= -\biggl({\bf e}_{\hat{{\bf n}}_2} \otimes
 {\bf e}_{\hat{{\bf n}}_2}- {\bf e}_{\hat{{\bf
 n}}_1} \otimes {\bf e}_{\hat{{\bf n}}_1}\biggr)=-{\bf e}_{+_{\hat{{\bf n}}_2\hat{{\bf
 n}}_2}}    \label{e71}   \\ & 
  {\bf e}_{\times_{\hat{{\bf n}}_1\hat{{\bf n}}_2}}= 
 {\bf e}_{\hat{{\bf n}}_1} \otimes {\bf e}_{\hat{{\bf n}}_2}+ {\bf e}_{\hat{{\bf
 n}}_2} \otimes {\bf e}_{\hat{{\bf n}}_1}= 
\biggl( {\bf e}_{\hat{{\bf n}}_2} \otimes {\bf e}_{\hat{{\bf n}}_1}+ {\bf e}_{\hat{{\bf
 n}}_1} \otimes {\bf e}_{\hat{{\bf n}}_2}\biggr)= {\bf e}_{\times_{\hat{{\bf
 n}}_2\hat{{\bf n}}_1}},    \nonumber \end{align} 
 The unit polarization tensors in the $(\hat{{\bf x}}, \ \hat{{\bf y}}, \
 \hat{{\bf z}})$ system may be  obtained from Eqs
 (\ref{e71}) by substituting , $\hat{{\bf n}}=\hat{{\bf
 z}}$, $\hat{{\bf n}}_1=\hat{{\bf
 x}}$,  $\hat{{\bf n}}_2=\hat{{\bf
 y}}$, \cite{mtw}
\begin{eqnarray} && {\bf e}_{+_{\hat{{\bf x}}\hat{{\bf x}}}}= 
 {\bf e}_{\hat{{\bf x}}} \otimes {\bf e}_{\hat{{\bf x}}}- {\bf e}_{\hat{{\bf
 y}}} \otimes {\bf e}_{\hat{{\bf y}}}= -\biggl({\bf e}_{\hat{{\bf y}}} \otimes
 {\bf e}_{\hat{{\bf y}}}- {\bf e}_{\hat{{\bf
 x}}} \otimes {\bf e}_{\hat{{\bf x}}}\biggr)=-{\bf e}_{+_{\hat{{\bf y}}\hat{{\bf
 y}}}}    \label{e72}   \\ && 
   {\bf e}_{\times_{\hat{{\bf x}}\hat{{\bf y}}}}= 
 {\bf e}_{\hat{{\bf x}}} \otimes {\bf e}_{\hat{{\bf y}}}+ {\bf e}_{\hat{{\bf
 y}}} \otimes {\bf e}_{\hat{{\bf x}}}= 
\biggl( {\bf e}_{\hat{{\bf y}}} \otimes {\bf e}_{\hat{{\bf x}}}+ {\bf e}_{\hat{{\bf
 x}}} \otimes {\bf e}_{\hat{{\bf y}}}\biggr)= {\bf e}_{\times_{\hat{{\bf
 y}}\hat{{\bf x}}}},    \nonumber \end{eqnarray} 
The corresponding unit polarization tensors in the cylindrical system 
$(\hat{{\bf
r}}, \hat{{\bf \phi}},
\hat{{\bf z}})$ for a  GW  which advances  in the $\hat{{\bf z}}$  
direction are detaily 
derived in \cite{bar2,bar3} from Eqs (\ref{e72}) by 
 using the transformation equations for  ${\bf e}_{\hat{{\bf x}}}$ 
 and ${\bf e}_{\hat{{\bf y}}}$ 
 \cite{spiegel}  
 \begin{equation} \label{e73}  {\bf e}_{\hat {\bf x}}=\cos(\phi){\bf e}_{\hat {\bf \rho}}-\sin(\phi){\bf
  e}_{\hat {\phi}}, \ \ {\bf e}_{\hat {\bf y}}=
  \sin(\phi){\bf e}_{\hat {\bf \rho}}+\cos(\phi){\bf
  e}_{\hat {\phi}}, \  \ {\bf e}_{\hat {\bf z}}= {\bf e}_{\hat {\bf z}}
  \end{equation}   
  That is,  using  the last Eqs (\ref{e73}) and the trigonometric identities  
  $(\cos^2(\phi)-\sin^2(\phi))=\cos(2\phi)$,  
 $2\cos(\phi)\sin(\phi)=\sin(2\phi)$,  one may obtain, as done in
 \cite{bar2,bar3},  the corresponding unit 
 polarization tensors in the 
 cylindrical system $(\hat{{\bf
r}}, \hat{{\bf \phi}},
\hat{{\bf z}})$,    denoted here 
   as 
${\bf e}^{(z)}_{+_{{\hat {\bf r}}{\hat {\bf r}}}} (=
-{\bf e}^{(z)}_{+_{{\hat {\bf \phi}}{\hat {\bf \phi}}}})$ and 
${\bf e}^{(z)}_{\times_{{\hat {\bf r}}{\hat {\bf \phi}}}} (=
{\bf e}^{(z)}_{\times_{{\hat {\bf \phi}}{\hat {\bf r}}}})$, 
\begin{align} & {\bf e}^{(z)}_{+_{{\hat {\bf r}}{\hat {\bf r}}}}=
 \cos(2\phi)\biggl( {\bf e}_{\hat {\bf r}}\otimes 
{\bf e}_{\hat {\bf r}}-{\bf e}_{\hat {\bf \phi}}\otimes 
{\bf e}_{\hat {\bf \phi}}\biggr)-
\sin(2\phi)\biggl( {\bf e}_{\hat {\bf r}}\otimes 
{\bf e}_{\hat {\bf \phi}}+{\bf e}_{\hat {\bf \phi}}\otimes 
{\bf e}_{\hat {\bf r}}\biggr) =- 
{\bf e}^{(z)}_{+_{{\hat {\bf \phi}}{\hat {\bf \phi}}}} \label{e74} \\ 
& {\bf e}^{(z)}_{\times_{{\hat {\bf r}}{\hat {\bf \phi}}}}=
 \sin(2\phi)\biggl( {\bf e}_{\hat {\bf r}}\otimes 
{\bf e}_{\hat {\bf r}}-{\bf e}_{\hat {\bf \phi}}\otimes 
{\bf e}_{\hat {\bf \phi}}\biggr)+\cos(2\phi)\biggl( {\bf e}_{\hat {\bf r}}\otimes 
{\bf e}_{\hat {\bf \phi}}+{\bf e}_{\hat {\bf \phi}}\otimes 
{\bf e}_{\hat {\bf r}}\biggr)={\bf e}^{(z)}_{\times_{{\hat {\bf \phi}}{\hat {\bf
r}}}}
 \nonumber \end{align} 
 Note  that since the two systems $(\hat{{\bf
r}}, \hat{{\bf \phi}},
\hat{{\bf z}})$ and $(\hat{{\bf
x}}, \hat{{\bf y}},
\hat{{\bf z}})$   
 are orthogonal the former unit polarization tensors 
${\bf e}^{(z)}_{+_{{\hat {\bf r}}{\hat {\bf r}}}}$,  
${\bf e}^{(z)}_{+_{{\hat {\bf \phi}}{\hat {\bf \phi}}}}$,   
${\bf e}^{(z)}_{\times_{{\hat {\bf r}}{\hat {\bf \phi}}}}$,  
${\bf e}^{(z)}_{\times_{{\hat {\bf \phi}}{\hat {\bf r}}}}$ 
  for a GW  advancing in the ${\bf z}$ direction of
 the $(\hat{{\bf
r}}, \hat{{\bf \phi}},
\hat{{\bf z}})$  system are derived here, as mentioned,  from the corresponding 
 unit polarization tensors 
${\bf e}_{+_{\hat{{\bf x}}\hat{{\bf x}}}}$, 
${\bf e}_{+_{\hat{{\bf y}}\hat{{\bf y}}}}$, 
${\bf e}_{+_{\hat{{\bf x}}\hat{{\bf y}}}}$, 
${\bf e}_{+_{\hat{{\bf y}}\hat{{\bf x}}}}$ of a GW
advancing in the ${\bf z}$ direction of
 the $(\hat{{\bf
x}}, \hat{{\bf y}},
\hat{{\bf z}})$ system. \par 
  We discuss here a cylindrical GW advancing in a general direction which may be
 decomposed along the  $(\hat{{\bf z}}, \ \hat{{\bf r}}, \hat{{\bf \phi}})$ 
 axes so
 that each of these components  may contribute its polarizing
 part to its respective perpendicular plane. Thus, the contribution to the unit
 polarization tensors in the  $(\hat{{\bf r}}, \ \hat{{ \bf \phi}})$ plane 
 resulting from the component along the 
 $\hat{{\bf z}}$ axis are given by Eqs (\ref{e74}). The contribution to the unit
 polarization tensors in the  $(\hat{{\bf r}}, \ \hat{{ \bf z}})$ plane 
 resulting  
 from the component along the 
 $\hat{{\bf \phi}}$ axis may be calculated in a similar manner, using Eqs
 (\ref{e73}),  as  
 \begin{align} & {\bf e}^{(\phi)}_{+_{{\hat {\bf r}}{\hat {\bf r}}}}=
 \cos^2(\phi)({\bf e}_{\hat {\bf r}}\otimes 
{\bf e}_{\hat {\bf r}})+\sin^2(\phi)({\bf e}_{\hat {\bf \phi}}\otimes 
{\bf e}_{\hat {\bf \phi}})-
\frac{1}{2}\sin(2\phi)\biggl( {\bf e}_{\hat {\bf r}}\otimes 
{\bf e}_{\hat {\bf \phi}}+{\bf e}_{\hat {\bf \phi}}\otimes 
{\bf e}_{\hat {\bf r}}\biggr)- \nonumber \\ & -
({\bf e}_{\hat 
{\bf z}}\otimes 
{\bf e}_{\hat {\bf z}}) = 
- {\bf e}^{(\phi)}_{+_{{\hat {\bf z}}{\hat {\bf z}}}} \label{e75} \\ 
& {\bf e}^{(\phi)}_{\times_{{\hat {\bf r}}{\hat {\bf z}}}}=
 \cos(\phi)\biggl( {\bf e}_{\hat {\bf r}}\otimes 
{\bf e}_{\hat {\bf z}}+{\bf e}_{\hat {\bf z}}\otimes 
{\bf e}_{\hat {\bf r}}\biggr)-\sin(\phi)\biggl( {\bf e}_{\hat {\bf \phi}}\otimes 
{\bf e}_{\hat {\bf z}}+{\bf e}_{\hat {\bf z}}\otimes 
{\bf e}_{\hat {\bf \phi}}\biggr)=
{\bf e}^{(\phi)}_{\times_{{\hat {\bf z}}{\hat {\bf r}}}}
 \nonumber \end{align} 
And the corresponding contribution to the unit
 polarization tensors in the  $(\hat{{\bf \phi}}, \ \hat{{ \bf z}})$ plane 
 resulting  
 from the component along the 
 $\hat{{\bf r}}$ axis is
\begin{align} & {\bf e}^{(r)}_{+_{{\hat {\bf \phi}}{\hat {\bf \phi}}}}=
 \cos^2(\phi)({\bf e}_{\hat {\bf \phi}}\otimes 
{\bf e}_{\hat {\bf \phi}})+\sin^2(\phi)({\bf e}_{\hat {\bf r}}\otimes 
{\bf e}_{\hat {\bf r}})+
\frac{1}{2}\sin(2\phi)\biggl( {\bf e}_{\hat {\bf r}}\otimes 
{\bf e}_{\hat {\bf \phi}}+{\bf e}_{\hat {\bf \phi}}\otimes 
{\bf e}_{\hat {\bf r}}\biggr)- \nonumber \\ & - 
({\bf e}_{\hat 
{\bf z}}\otimes 
{\bf e}_{\hat {\bf z}})=  - 
{\bf e}^{(r)}_{+_{{\hat {\bf z}}{\hat {\bf z}}}}
   \label{e76} \\ 
& {\bf e}^{(r)}_{\times_{{\hat {\bf \phi}}{\hat {\bf z}}}}=
 \cos(\phi)\biggl( {\bf e}_{\hat {\bf \phi}}\otimes 
{\bf e}_{\hat {\bf z}}+{\bf e}_{\hat {\bf z}}\otimes 
{\bf e}_{\hat {\bf \phi}}\biggr)+\sin(\phi)\biggl( {\bf e}_{\hat {\bf r}}\otimes 
{\bf e}_{\hat {\bf z}}+{\bf e}_{\hat {\bf z}} \otimes 
{\bf e}_{\hat {\bf r}}\biggr)=
{\bf e}^{(r)}_{\times_{{\hat {\bf z}}{\hat {\bf \phi}}}}
 \nonumber \end{align} 
Thus, the total unit polarization tensor in the $\hat{{\bf r}}\hat{{\bf r}}$
direction may
be found from Eqs (\ref{e74})-(\ref{e75}) as 
\begin{align} & {\bf e}^{(total)}_{+_{{\hat {\bf r}}{\hat {\bf r}}}}=
{\bf e}^{(z)}_{+_{{\hat {\bf r}}{\hat {\bf r}}}}+
{\bf e}^{(\phi)}_{+_{{\hat {\bf r}}{\hat {\bf r}}}}=
\biggl(\cos(2\phi)+\cos^2(\phi)\biggr)({\bf e}_{\hat {\bf r}}\otimes 
{\bf e}_{\hat {\bf r}})
+\biggl(\sin^2(\phi)-\cos(2\phi)\biggr) \cdot \nonumber \\ & \cdot 
({\bf e}_{\hat {\bf \phi}}\otimes 
{\bf e}_{\hat {\bf \phi}}) -\frac{3}{2}\sin(2\phi)\biggl( {\bf e}_{\hat {\bf r}}\otimes 
{\bf e}_{\hat {\bf \phi}}+{\bf e}_{\hat {\bf \phi}}\otimes 
{\bf e}_{\hat {\bf r}}\biggr)-
({\bf e}_{\hat 
{\bf z}}\otimes 
{\bf e}_{\hat {\bf z}})
\label{e77} \end{align} 
And the total unit polarization tensor in the $\hat{{\bf \phi}}\hat{{\bf \phi}}$
direction may, analogously, be found  from Eqs (\ref{e74}) and (\ref{e76}) as 
\begin{align} & {\bf e}^{(total)}_{+_{{\hat {\bf \phi}}{\hat {\bf \phi}}}}=
{\bf e}^{(z)}_{+_{{\hat {\bf \phi}}{\hat {\bf \phi}}}}+
{\bf e}^{(r)}_{+_{{\hat {\bf \phi}}{\hat {\bf \phi}}}}=
\biggl(\sin^2(\phi)-\cos(2\phi)\biggr)({\bf e}_{\hat {\bf r}}\otimes 
{\bf e}_{\hat {\bf r}})
+\biggl(\cos^2(\phi)+\cos(2\phi)\biggr) \cdot \nonumber \\ & \cdot 
({\bf e}_{\hat {\bf \phi}}\otimes 
{\bf e}_{\hat {\bf \phi}}) +\frac{3}{2}\sin(2\phi)\biggl( {\bf e}_{\hat 
{\bf \rho}}\otimes 
{\bf e}_{\hat {\bf \phi}}+{\bf e}_{\hat {\bf \phi}}\otimes 
{\bf e}_{\hat {\bf r}}\biggr)-
({\bf e}_{\hat 
{\bf z}}\otimes 
{\bf e}_{\hat {\bf z}})
\label{e78} \end{align} 
Likewise,  the total unit polarization tensor in the $\hat{{\bf z}}\hat{{\bf 
z}}$
direction may be found  from Eqs (\ref{e75})-(\ref{e76}) as 
\begin{equation}   \label{e79}  
{\bf e}^{(total)}_{+_{{\hat {\bf z}}{\hat {\bf z}}}}=
{\bf e}^{(r)}_{+_{{\hat {\bf z}}{\hat {\bf z}}}}+
{\bf e}^{(\phi)}_{+_{{\hat {\bf z}}{\hat {\bf z}}}}=
2({\bf e}_{\hat 
{\bf z}}\otimes 
{\bf e}_{\hat {\bf z}})-\biggl({\bf e}_{\hat 
{\bf r}}\otimes 
{\bf e}_{\hat {\bf r}}+{\bf e}_{\hat {\bf \phi}}\otimes 
{\bf e}_{\hat {\bf \phi}}\biggr) \end{equation}
As for the polarization tensors in the directions $\hat{{\bf j}}\hat{{\bf 
k}}$  where $\hat{{\bf j}} \neq \hat{{\bf k}}$ these should certainly 
 be
generated by the mixed GW's $h^{TT}_{jk_{j\neq k}}$ but as seen from Eqs
(\ref{e63}) and (\ref{e68})-(\ref{e70}) no such cylindrical GW's exist in
either the TT gauge or in the linearized version introduced by Eqs
(\ref{e61})-(\ref{e64}). 
Thus, the general cylindrical GW, denoted here  $h^{TT}_{(+,\times)_{jk}}$, 
may be simplified as 
\begin{equation} h^{TT}_{(+,\times)_{jk}}= h^{TT}_{+_{jk}}+
h^{TT}_{\times_{jk}}=  \label{e80} 
 A_{+_{jk}}{\bf e}^{(total)}_{+_{\hat{{\bf jk}}}}h^{TT}_{jk_{j=k}}+
A_{\times_{jk}}{\bf e}^{(total)}_{\times_{\hat{{\bf jk}}}}h^{TT}_{jk_{j\neq k}}=
A_{+_{jk}}{\bf e}^{(total)}_{+_{\hat{{\bf jk}}}}h^{TT}_{jk_{j=k}},
\end{equation} 
where $A_{+_{jk}}$ and $A_{\times_{jk}}$ are constants \cite{mtw,bar2,bar3}  
related,  respectively, 
 to $h^{TT}_{jk_{j=k}}$ and $h^{TT}_{jk_{j\neq k}}$ and 
 ${\bf e}^{(total)}_{+_{\hat{{\bf jk}}}}$ are given by Eqs
 (\ref{e77})-(\ref{e79}).  Substituting
in Eq (\ref{e80}) 
from Eqs (\ref{e68})-(\ref{e70}) for $h^{TT}_{jk_{j=k}}$ and from 
 Eqs (\ref{e77})-(\ref{e79}) for ${\bf e}^{(total)}_{+_{\hat{{\bf jk}}}}$ one obtains 
 \begin{align} & h^{TT}_{+_{rr}}=
 A_{+_{rr}}{\bf e}^{(total)}_{+_{\hat{{\bf rr}}}}h^{TT}_{rr}=A_{+_{rr}}\biggl\{\biggl(\cos(2\phi)+
 \cos^2(\phi)\biggr)({\bf e}_{\hat {\bf \rho}}\otimes 
{\bf e}_{\hat {\bf r}})
+\biggl(\sin^2(\phi)- \nonumber \\ & - \cos(2\phi)\biggr)  
({\bf e}_{\hat {\bf \phi}}\otimes 
{\bf e}_{\hat {\bf \phi}}) -\frac{3}{2}\sin(2\phi)\biggl( {\bf e}_{\hat {\bf r}}\otimes 
{\bf e}_{\hat {\bf \phi}}+{\bf e}_{\hat {\bf \phi}}\otimes 
{\bf e}_{\hat {\bf r}}\biggr)-
({\bf e}_{\hat 
{\bf z}}\otimes 
{\bf e}_{\hat {\bf z}})\biggr\}\cdot \label{e81} \\ & \cdot 
\biggl\{c_1+c_2t-\frac{1}{2}\biggl[\int_0^{\infty}dk
 \biggl(\bigl(J_0(kr)+J_2(kr)\bigr) \cdot 
 \cos(kT)\bigl(A(k)+A^+(k)\bigr)- \nonumber \\ & -
4kA(k)A^+(k)\biggr)\biggr] \biggr\}
\nonumber \end{align}
 \begin{align} & h^{TT}_{+_{\phi\phi}}=
  A_{+_{\phi\phi}}{\bf e}^{(total)}_{+_{\hat{{\bf \phi\phi}}}}h^{TT}_{\phi\phi}=
  A_{+_{\phi\phi}}\biggl\{
 \biggl(\sin^2(\phi)-\cos(2\phi)\biggr)({\bf e}_{\hat {\bf r}}\otimes 
{\bf e}_{\hat {\bf r}})
+\biggl(\cos^2(\phi)+ \nonumber \\ & + 
\cos(2\phi)\biggr)  
({\bf e}_{\hat {\bf \phi}}\otimes 
{\bf e}_{\hat {\bf \phi}}) +\frac{3}{2}\sin(2\phi)\biggl( {\bf e}_{\hat 
{\bf r}}\otimes 
{\bf e}_{\hat {\bf \phi}}+{\bf e}_{\hat {\bf \phi}}\otimes 
{\bf e}_{\hat {\bf r}}\biggr)-
({\bf e}_{\hat 
{\bf z}}\otimes 
{\bf e}_{\hat {\bf z}})\biggr\} \cdot \label{e82}  \\ & 
\cdot \biggl\{c_1+c_2t+\ln(R^2)-
\biggl(\int_0^{\infty}J_0(kr)\cos(kT)(A(k)+A^+(k))\biggr)\biggr\}
\nonumber \end{align}
\begin{align} & h^{TT}_{+_{zz}}= A_{+_{zz}}{\bf e}^{(total)}_{+_{\hat{{\bf
zz}}}}h^{TT}_{zz}= A_{+_{zz}}\biggl\{2({\bf e}_{\hat 
{\bf z}}\otimes 
{\bf e}_{\hat {\bf z}})-\biggl({\bf e}_{\hat 
{\bf r}}\otimes 
{\bf e}_{\hat {\bf r}}+{\bf e}_{\hat {\bf \phi}}\otimes 
{\bf e}_{\hat {\bf \phi}}\biggr)\biggr\} \cdot \label{e83} \\  & \cdot 
\biggl\{c_1+c_2t+\int_0^{\infty}J_0(kr)\cos(kT)(A(k)+A^+(k)) \biggr\}
\nonumber   \end{align}
The corresponding cylindrical metric in the TT gauge may, now, be written as 
\begin{equation} \label{e84}  
(ds^{TT})_{({\hat{\bf r}},{\hat{\bf \phi}},
{\hat{\bf z}})}^2=
h^{TT}_{+_{rr}}dr^2+
h^{TT}_{+_{\phi\phi}}d\phi^2+
h^{TT}_{+_{zz}}dz^2, \end{equation} 
where $h^{TT}_{+_{rr}}$, $h^{TT}_{+_{\phi\phi}}$ and $h^{TT}_{+_{zz}}$ are, 
respectively, given by Eqs  (\ref{e81})-(\ref{e83}).

\markright{THE EMBEDDED TRAPPED SURFACE GENERATED BY THE CYLINDRICAL....}

\protect \section{The embedded trapped surface generated by the cylindrical gravitational
wave}

We first note in the context of   trapped surfaces that they are, generaly, 
 generated by very strong gravitational waves \cite{mtw,eppley} and not by the
 weak ones discussed in the former sections. But, as emphasized in
 \cite{bar2,bar3}, if these weak GW's persist somehow in some spacetime
 region for a long time then these GW's  are added and aggregated \cite{mtw} 
 upon each other 
 in such a way that they  may influence a spacetime region as if a very
 strong GW passes this region in a short time. Thus, we discuss here, 
 as in
 \cite{bar2,bar3}, GW which dwell in some spacetime region enough time 
 to generate trapped surface. \par 
We, now, calculate and find the embedded trapped surface generated by the 
cylindrical GW. We note that since it is difficult to embed the whole trapped
surface \cite{eppley} one resorts to the simpler task of embedding the
equatorial plane due to its rotational symmetry. We use for that the method 
in \cite{eppley} and begin by requiring the metric on the equator to be equal to
that of a surface of rotation $z(x,y)$ in Euclidean space. 
\begin{equation} x=F(r)\cos(\phi), \ \ y=F(r)\sin(\phi), \ \ z=G(r) 
 \label{e85} \end{equation} 
 Thus, taking into account that on the equator $d^2z=0$ we obtain for the
 metrics
\begin{align}  &   
 ds^2=dx^2+dy^2+dz^2=\biggl(F_{,\rho}(r)^2+G_{,\rho}(r)^2\biggr)dr^2+
 F^2(r)d\phi^2=  \nonumber \\ & = 
 h^{TT}_{+_{rr}}dr^2 +
h^{TT}_{+_{\phi\phi}}d^2\phi  =  
A_{+_{rr}}{\bf e}^{(total)}_{+_{\hat{{\bf rr}}}}h^{TT}_{rr}d^2r+
A_{+_{\phi\phi}}{\bf e}^{(total)}_{+_{\hat{{\bf
\phi\phi}}}}h^{TT}_{\phi\phi}d^2\phi= \nonumber \\ & =
A_{+_{rr}}\biggl\{\biggl(\cos(2\phi)+
 \cos^2(\phi)\biggr)({\bf e}_{\hat {\bf \rho}}\otimes 
{\bf e}_{\hat {\bf r}})
+\biggl(\sin^2(\phi)- \cos(2\phi)\biggr) \cdot \nonumber \\ & \cdot 
({\bf e}_{\hat {\bf \phi}}\otimes 
{\bf e}_{\hat {\bf \phi}}) -\frac{3}{2}\sin(2\phi)\biggl( {\bf e}_{\hat {\bf r}}\otimes 
{\bf e}_{\hat {\bf \phi}}+{\bf e}_{\hat {\bf \phi}}\otimes 
{\bf e}_{\hat {\bf r}}\biggr)-
({\bf e}_{\hat 
{\bf z}}\otimes 
{\bf e}_{\hat {\bf z}})\biggr\}\cdot \label{e86} \\ & 
\cdot \biggl\{c_1+c_2t-\frac{1}{2}\biggl[\int_0^{\infty}dk
 \biggl(\bigl(J_0(kr)+J_2(kr)\bigr) \cdot 
 \cos(kT)\bigl(A(k)+A^+(k)\bigr)- \nonumber \\ & -
4kA(k)A^+(k)\biggr)\biggr] \biggr\}d^2r+ 
  A_{+_{\phi\phi}}\biggl\{
 \biggl(\sin^2(\phi)-\cos(2\phi)\biggr)({\bf e}_{\hat {\bf r}}\otimes 
{\bf e}_{\hat {\bf r}})
+\nonumber \\ & +\biggl(\cos^2(\phi)+ 
\cos(2\phi)\biggr)  
({\bf e}_{\hat {\bf \phi}}\otimes 
{\bf e}_{\hat {\bf \phi}}) +\frac{3}{2}\sin(2\phi)\biggl( {\bf e}_{\hat 
{\bf r}}\otimes 
{\bf e}_{\hat {\bf \phi}}+{\bf e}_{\hat {\bf \phi}}\otimes 
{\bf e}_{\hat {\bf r}}\biggr)- \nonumber \\ & - 
({\bf e}_{\hat 
{\bf z}}\otimes 
{\bf e}_{\hat {\bf z}})\biggr\} 
 \biggl\{c_1+c_2t+\ln(R^2)-
\biggl(\int_0^{\infty}dkJ_0(kr)\cos(kT)(A(k)+A^+(k))\biggr)\biggr\}d^2\phi
\nonumber
\end{align}
From the last equation one obtains for the quantities $F(r)$, $F_{,r}(r)$ and 
$G(r)$
\begin{align} & F(r)= A_{+_{\phi\phi}}\biggl\{
 \biggl(\sin^2(\phi)-\cos(2\phi)\biggr)({\bf e}_{\hat {\bf r}}\otimes 
{\bf e}_{\hat {\bf r}})
+\biggl(\cos^2(\phi)+ 
\cos(2\phi)\biggr) \cdot \nonumber \\ &  \cdot 
({\bf e}_{\hat {\bf \phi}}\otimes 
{\bf e}_{\hat {\bf \phi}}) +\frac{3}{2}\sin(2\phi)\biggl( {\bf e}_{\hat 
{\bf r}}\otimes 
{\bf e}_{\hat {\bf \phi}}+{\bf e}_{\hat {\bf \phi}}\otimes 
{\bf e}_{\hat {\bf r}}\biggr) - 
({\bf e}_{\hat 
{\bf z}}\otimes 
{\bf e}_{\hat {\bf z}})\biggr\} \cdot \label{e87} \\ & \cdot
 \biggl\{c_1+c_2t+\ln(R^2)-
\biggl(\int_0^{\infty}dkJ_0(kr)\cos(kT)(A(k)+A^+(k))\biggr)\biggr\}
\nonumber \end{align} 
\begin{align} & F_{,r}(r)=A_{+_{\phi\phi}} \biggl(\frac{2}{R}+
\biggl(\int_0^{\infty}dkkJ_1(kr)\cos(kT)(A(k)+A^+(k))\biggr)\biggr)
\cdot \nonumber \\ & \cdot
\biggl\{
 \biggl(\sin^2(\phi)- 
 \cos(2\phi)\biggr)({\bf e}_{\hat {\bf r}}\otimes 
{\bf e}_{\hat {\bf r}})
+\biggl(\cos^2(\phi)+ 
\cos(2\phi)\biggr) \cdot \label{e88} \\ &  \cdot 
({\bf e}_{\hat {\bf \phi}}\otimes 
{\bf e}_{\hat {\bf \phi}}) +\frac{3}{2}\sin(2\phi)\biggl( {\bf e}_{\hat 
{\bf r}}\otimes 
{\bf e}_{\hat {\bf \phi}}+{\bf e}_{\hat {\bf \phi}}\otimes 
{\bf e}_{\hat {\bf r}}\biggr) - 
({\bf e}_{\hat 
{\bf z}}\otimes 
{\bf e}_{\hat {\bf z}})\biggr\}, 
\nonumber \end{align}
where the Bessel's derivative $\frac{dJ_0(r)}{dr}=-J_1(r)$ is used.
\begin{align} & G(r)=\int dr\biggl\{A_{+_{rr}}\biggl\{\biggl(\cos(2\phi)+
 \cos^2(\phi)\biggr)({\bf e}_{\hat {\bf \rho}}\otimes 
{\bf e}_{\hat {\bf r}})
+\biggl(\sin^2(\phi)- \cos(2\phi)\biggr) \cdot \nonumber \\ & \cdot 
({\bf e}_{\hat {\bf \phi}}\otimes 
{\bf e}_{\hat {\bf \phi}}) -\frac{3}{2}\sin(2\phi)\biggl( {\bf e}_{\hat {\bf r}}\otimes 
{\bf e}_{\hat {\bf \phi}}+{\bf e}_{\hat {\bf \phi}}\otimes 
{\bf e}_{\hat {\bf r}}\biggr)-
({\bf e}_{\hat 
{\bf z}}\otimes 
{\bf e}_{\hat {\bf z}})\biggr\}\cdot \label{e89} \\ & 
\cdot \biggl\{c_1+c_2t-\frac{1}{2}\biggl[\int_0^{\infty}dk
 \biggl(\bigl(J_0(kr)+J_2(kr)\bigr) \cdot 
 \cos(kT)\bigl(A(k)+A^+(k)\bigr)- \nonumber \\ & -
4kA(k)A^+(k)\biggr)\biggr] \biggr\}-F_{,r}^2(r)\biggr\}^{\frac{1}{2}},
\nonumber \end{align}
where $F_{,r}(r)$ is given by Eq (\ref{e88}). The expressions $F(r)$, 
$F_{,r}(r)$ and
$G(r)$ determine the geometry of the cylindrical trapped surface.

\markright{CONCLUDING REMARKS} 

\protect \section{Concluding Remarks}

We have discussed quantum aspects of the cylindrical  GW and, especially, the
theoretical possibility of increasing its quantum probability.  
We use for that  the  spatial version \cite{bar1} of the Zeno effect
\cite{zeno,aharonov,facchi} which is affected 
  by   performing  the same experiment in a large number of
nonoverlapping separate regions of space all included in a finite total region
so that in the limit in which these subregions become infinitesimal, keeping the
total region fixed,  the quantum
state becomes constant in space. 
The last spatial method of the Zeno effect is more appropriate for
the cylindrical GW  since it  is related  \cite{kuchar1} to the 
extrinsic time  variable which  
 is canonically conjugate to momentum \cite{kuchar1} just as the
spatial coordinate is canonicaly conjugate to it. Thus, we have shown that,
beginning with some cylindrical GW  in some certain subregion of space-time
included in a larger one and if the measurement  of this field is similarly done
in other neighbouring space-time subregions   then at
the limit in which these subregions become infinitesimally small, keeping the
including larger one fixed,  one obtains the
Zeno result in which the cylindrical GW  is fixed in all these subregions. 
For that  we have first detaily shown that  the cylindrical GW   may  be
appropriately discussed in the 
  commutation number representation \cite{schiff,merzbacher}. 
  In this representation the cylindrical GW is shown to be 
  composed   from a
  large ensemble of particle-like components each one inhabits some 
 space-time  point   so that it may  be discussed in terms of space Zeno
  effect which also requires a large number of similar components 
   to dwell in similar points all confined in some
  finite region of space-time. \par
   As known \cite{mtw}, in the canonical formulation of general relativity 
   the lapse and
  shift functions,  which are fixed  (by observer) 
   upon some space-time  hypersurface,   determine \cite{mtw} the later 
   evolution and geometry 
   of space-time.  This is reminiscent
  of the similar role of the  creation and destruction operators in the
  commutation number representation which also determine space-time geometry 
   through controlling the number of gravitational constituents of the GW  
  which impose its geometry upon space-time.   In Section IV and Appendix 
  $A$ we have
  expressed these lapse and shift functions in terms of the creation and
  destruction operators. Moreover, we were not contented in only finding 
     the conditions through which the probability of the GW  increases but
  also follow the "realized"  (with unity-value  probability)   GW  
  in its passage through space-time. We have, thus,  
  discussed  its properties in the (TT) gauge and have calculated 
   the trapped surface 
   generated by it. \par
  It must be noted  that although the time version of the 
  quantum Zeno effect
  was experimentally validated \cite{itano,kofman}  no such corroboration 
   exists, for
  now,  
  for its spatial version \cite{bar1} and less for the gravitational 
  application
  of it that we use here. This is because   the GW   itself were
  not experimentally, up to now,   detected.  
  One  hope that
  some future technology will detect not only   GW  but also a possible Zeno
  effect for them. That is, one may hope that as this effect proves itself so
  efficiently in the quantum regime it may, likewise, also be efficient in
  fixing and classicalizing    GW  and by that  fixing 
   its imposed geometry upon the surrounding spacetime.   

\markright{APPENDIX A}

\appendix 

\section{ Calculation of the expressions $(C_1C_4+C_2C_3)$, 
$(C_1C_2-C_4C_3)$, $(C_1C_1+C_3C_3)$ from Eq (\ref{e44})}

We, now, calculate and simplify the expressions $(C_1C_4+C_2C_3)$, 
$(C_1C_2-C_4C_3)$, $(C_1C_1+C_3C_3)$. We begin with $(C_1C_4+C_2C_3)$ and 
use the commutation relations (\ref{e18}) for $A(k)$, the trigonometric
identities: $\sin(a)\sin(b)=\frac{1}{2}\bigl(\cos(a-b)-\cos(a+b)\bigr)$, 
 $\cos(a)\cos(b)=\frac{1}{2}\bigl(\cos(a-b)+\cos(a+b)\bigr)$ and the 
  expression (see integral 11.4.42 in P. 487 in \cite{abramowitz})
\begin{equation} \int_0^{\infty}J_{\mu}(ar)J_{\mu-1}(br)dr=   \tag{$A_1$}
\label{$A_1$}
\left\{ \begin{array}{ll} \frac{b^{(\mu-1)}}{a^{\mu}} &  {\rm for (0 < b < a) \
and \ \Re(\mu) > 0} 
\\ \frac{1}{2b} & 
 {\rm for (0 < b = a) \ and \ \Re(\mu) > 0 } \\ 0  & 
 {\rm for (b > a > 0 \  and \  \Re(\mu) > 0} \end{array} \right.
\end{equation}
 Thus, the expression $(C_1C_4+C_2C_3)$ is 
 \begin{align} &
C_1C_4+C_2C_3=\int_0^{\infty}dr\int_0^{\infty}dk\int_0^{\infty}dk'
kk'J_0(kr)J_1(k'r)\biggl\{\sin(kT)\sin(k'T) \cdot \nonumber \\ &
\cdot \biggl(A(k)A(k')-A(k)A^+(k')+A^+(k)A(k')-A^+(k)A^+(k')\biggr)+
\cos(kT)\cos(k'T) \cdot \nonumber \\ & \cdot 
\biggl(A(k')A(k)-A(k')A^+(k)+A^+(k')A(k)-A^+(k')A^+(k)\biggr)\biggr\}= \nonumber
\\ & = \tag{$A_2$} \label{$A_2$}
\int_0^{\infty}dr\int_0^{\infty}dk\int_0^{\infty}dk'
kk'J_0(kr)J_1(k'r)\biggl\{\frac{\cos(T(k-k'))}{2}
\biggl(2A(k)A(k')-  \nonumber \\ & - 2A^+(k)A^+(k')
 -\frac{\delta(k-k')}{2}-\frac{\delta(k'-k)}{2}\biggr)
+\frac{1}{2}\cos(T(k+k')) 
\biggl(A^+(k')A(k)+ \nonumber \\ & + A(k)A^+(k')-A^+(k)A(k')-
A(k')A^+(k)\biggr)\biggr\}=\int_0^{\infty}dr\int_0^{\infty}dk\int_0^{\infty}dk'
kk' \nonumber \\ & \cdot J_0(kr)J_1(k'r)  
\biggl\{\frac{\cos(T(k-k'))}{2} 
\biggl(2A(k)A(k')-
2A^+(k)A^+(k') -1 \biggr) + \nonumber \\ & + 
\frac{\cos(T(k+k'))}{2}\biggl(2A(k)A^+(k')-2A(k')A^+(k)-
\frac{\delta(k-k')}{2}+
\frac{\delta(k'-k)}{2} \biggr\}= \nonumber \\ & =
\frac{1}{2}\int_0^{\infty}dkk\biggl(A(k)A(k)-A^+(k)A^+(k)-\frac{1}{2}\biggr) 
\nonumber \end{align}
Note that in the last result we have equated $k$ to $k'$ and use the middle
relation from Eqs (\ref{$A_1$}) otherwise if we assume $k \neq k'$ we would get 
either  an imaginary result for   
 $(C_1C_4+C_2C_3)$  if we use  the first  relation of (\ref{$A_1$}) or 
 $(C_1C_4+C_2C_3)=0$ if we use  the third  relation of it. These outcomes   
  would make $N^0$ and $N^1$ from Eqs (\ref{e44}) either imaginary because of
  the first of (\ref{$A_1$}) or undefined because of the third of it. 
. \par 
In a similar manner, using 
the commutation relations (\ref{e18}) for $A(k)$, the trigonometric
identities $\sin(a)\cos(b)=\frac{1}{2}\bigl(\sin(a-b)+\sin(a+b)\bigr)$,
$\sin(a-b)=-\sin(b-a)$   
  and the integral (\ref{$A_1$}),   the expression $C_1C_2-C_4C_3$ may be 
  calculated as \begin{align} & C_1C_2-C_4C_3=\int_0^{\infty}dr\int_0^{\infty}dk\int_0^{\infty}dk'
kk'J_0(kr)J_1(k'r)\biggl\{\sin(kT)\cos(k'T) \cdot \nonumber \\ &
\cdot \biggl(A(k)A(k')+A(k)A^+(k')+A^+(k)A(k')+A^+(k)A^+(k')\biggr)-
\sin(k'T)\cos(kT) \cdot \nonumber \\ & \cdot 
\biggl(A(k')A(k)-A(k')A^+(k)-A^+(k')A(k)+A^+(k')A^+(k)\biggr)\biggr\}= 
\tag{$A_3$} \label{$A_3$}
\\ & = \int_0^{\infty}dr\int_0^{\infty}dk\int_0^{\infty}dk'
kk'J_0(kr)J_1(k'r)\biggl\{\frac{\sin(T(k+k'))}{2}\biggl(A(k)A^+(k')+ \nonumber
\\ & +
A^+(k)A(k')+A(k')A^+(k)+A^+(k')A(k)\biggr) 
+\frac{\sin(T(k-k'))}{2}\biggl(2A(k)A(k')+ \nonumber \\ & +2A^+(k)A^+(k')
+A(k)A^+(k')+A^+(k)A(k')-A(k')A^+(k)-A^+(k')A(k) \biggr)\biggr\}= \nonumber \\
& =
\int_0^{\infty}dr\int_0^{\infty}dk\int_0^{\infty}dk'
kk'J_0(kr)J_1(k'r)\biggl\{\frac{\sin(T(k+k'))}{2}
\biggl(2A(k)A^+(k')+ \nonumber \\ & + 2A(k')A^+(k)
 -\frac{\delta(k-k')}{2}-\frac{\delta(k'-k)}{2}\biggr)+\frac{\sin(T(k-k'))}{2}
\biggl(2A(k)A(k')+ \nonumber \\ & + 2A^+(k)A^+(k')
 +\frac{\delta(k-k')}{2}-\frac{\delta(k'-k)}{2}\biggr)\biggr\}=
 \frac{1}{2}\int_0^{\infty}dkk\sin(2Tk)\biggl(2A(k)A^+(k')-\frac{1}{2}\biggr)
\nonumber
\end{align} 
As realized from the last result we have assumed $k=k'$ otherwise, as noted
after Eq (\ref{$A_2$}), we   would obtain either an imaginary or zero value for 
$(C_1C_4+C_2C_3)$. 
These would make  $N^0$ and $N^1$, as noted
after Eq (\ref{$A_2$}), either imaginary or undefined. 
 \par
The remaining expression $C_1C_1+C_3C_3$ is calculated by using the former
product trigonometric identities and the 
orthogonality relation (see the unnumbered expression after Eq (\ref{e19})) 
$\int_0^{\infty}dkkJ_n(kr)J_n(kr')=\frac{\delta(r-r')}{r'}$
\begin{align} & C_1C_1+C_3C_3=\int_0^{\infty}dr\int_0^{\infty}dk\int_0^{\infty}dk'
kk'J_0(kr)J_0(k'r)\biggl\{\sin(kT)\sin(k'T) \cdot \nonumber \\ &
\cdot \biggl(A(k)A(k')+A(k)A^+(k')+A^+(k)A(k')+A^+(k)A^+(k')\biggr)+
\cos(kT)\cos(k'T) \cdot \nonumber \\ & \cdot 
\biggl(A(k')A(k)-A(k')A^+(k)-A^+(k')A(k)+A^+(k')A^+(k)\biggr)\biggr\}= 
\tag{$A_4$} \label{$A_4$}
\\ & =
\int_0^{\infty}dr\int_0^{\infty}dk\int_0^{\infty}dk'
kk'J_0(kr)J_0(k'r)\biggl\{\frac{\cos(T(k-k'))}{2}
\biggl(2A(k)A^+(k')-  \nonumber \\ & + 2A(k')A^+(k)
 -\frac{\delta(k-k')}{2}-\frac{\delta(k'-k)}{2}\biggr)
-\frac{1}{2}\cos(T(k+k')) 
\biggl(2A(k)A(k')+ \nonumber \\ & + 2A^+(k)A^+(k')
+\frac{\delta(k-k')}{2}-\frac{\delta(k'-k)}{2}\biggr)
\biggr)\biggr\}=\nonumber \\ & = \int_0^{\infty}dr\int_0^{\infty}dk\frac{\delta(r-r)}{r}
kk  \biggl\{\frac{1}{2} 
\bigl(4A(k)A(k) -1 \bigr) -
\cos(2Tk)\biggl(A(k)A(k)+ \nonumber \\ & +A^+(k)A^+(k)
\biggr\}= 
\int_0^{\infty}dk\frac{k^2}{r}\biggl\{\biggl(2A(k)A^+(k)-\frac{1}{2}\biggr)
-\cos(2Tk)\biggl(A(k)A(k)+A^+(k)A^+(k)\biggr)\biggr\} 
\nonumber \end{align}

\markright{APPENDIX B}

\protect\section{   Derivation of the probability 
$Pr^{(1)}(|\Psi(0,\psi(R))\!>,\rho)$ from  Eq (\ref{e51})}
We use 
 the shift operator \cite{schiff,merzbacher} $ e^{\frac{i\rho P}{\hbar}}$  
 where $P$ denote the momentum
 and $\rho$ is, as mentioned,  a very small amount by which the state 
  functional shifts
 \cite{schiff,merzbacher}. Note, as mentioned, that the extrinsic time variable
 related to the cylindrical   GW is, like any spatial variable, canonically
 conjugate \cite{kuchar1} to momentum.   Thus,  
   one may write for this
 probability $Pr^{(1)}(|\Psi(0,\psi(R))\!>,\rho)$

\begin{align}  &  Pr^{(1)}(|\Psi(0,\psi(R))\!>,\rho)=
 \biggl|<\!\Psi(T(R),\psi(R))\biggl|e^{\frac{i\rho
P}{\hbar}}\biggr|\Psi(T(R),\psi(R))\!>\biggr|^2 =\biggl| <\!\Psi(0,\psi(R))|
\cdot \nonumber \\ & \cdot 
\exp\biggl\{i\biggl[\frac{T(R)}{2(1-T^2_{,R}(R))} \biggl(R^{-1}\pi^2_{\psi}(R)
-2T_{,R}(R)\pi_{\psi}(R)\psi_{,R}(R) + 
 \frac{1}{2}R\psi^2_{,R}(R)\biggr) \biggr] \biggr\} \cdot \nonumber \\ & \cdot 
 \biggl|e^{\frac{i\rho
P}{\hbar}}\biggr|
\exp\biggl\{-i\biggr[\frac{T(R)}{2(1-T^2_{,R}(R))}\biggl(R^{-1}\pi^2_{\psi}(R)
-2T_{,R}(R)\pi_{\psi}(R)\psi_{,R}(R)  + \nonumber \\ & + \frac{1}{2}R\psi^2_{,R}(R)
\biggr)\biggr]\biggr\}  |\Psi(0,\psi(R))\!>\biggr|^2
 =\biggl| <\!\Psi(0,\psi(R))| 
  \exp\biggl\{i\biggl[\frac{T(R)}{2R(1-T^2_{,R}(R))} 
   \tag{$B_1$} \label{$B_1$}  \\ & \cdot  \biggl(\pi^2_{\psi}(R) 
-2RT_{,R}(R)\pi_{\psi}(R)\psi_{,R}(R) + 
\frac{1}{2}R^2\psi^2_{,R}(R)\biggr) \biggr] \biggr\}
\exp\biggl\{-i \nonumber  \\ & \cdot 
\biggr[\frac{T(R+\rho)}{2(R+\rho)\biggl(1-T^2_{,R}(R+\rho)\biggr)}  
\biggl(
\pi^2_{\psi}(R+\rho)  
-2(R+\rho)T_{,R}(R+\rho) \nonumber \\ & \cdot \pi_{\psi}(R+\rho)  
\psi_{,R}(R+\rho) +  
 \frac{1}{2}(R+\rho)^2\psi^2_{,R}(R+\rho)
\biggr)\biggr]\biggr\}|\Psi(0,\psi(R+ \rho))\!>\biggr|^2 \nonumber
\end{align} 
Taking common denominator for the exponential expressions one obtains from the
last result
\begin{align}  &  Pr^{(1)}(|\Psi(0,\psi(R))\!>,\rho)=
\biggl| <\!\Psi(0,\psi(R))|\exp\biggl\{i\biggl[\frac{1}{2R(R+\rho)
(1-T^2_{,R}(R))\biggl(1-T^2_{,R}(R+\rho)\biggr)}  \nonumber \\ & \cdot 
\biggl\{T(R)\biggl[(R+\rho)\biggl(1-T^2_{,R}(R+\rho)\biggr)
\pi^2_{\psi}(R)
-2R(R+\rho)\biggl(1-T^2_{,R}(R+\rho)\biggr)\cdot \nonumber \\ & \cdot 
T_{,R}(R)\pi_{\psi}(R)\psi_{,R}(R) 
 +\frac{1}{2}R^2(R+\rho)\biggl(1-T^2_{,R}(R+\rho)\biggr) \psi^2_{,R}(R)\biggr]- 
\tag{$B_2$} \label{$B_2$} \\ & 
- T(R+\rho)\biggl[R\biggl(1-T^2_{,R}(R)\biggr)\pi^2_{\psi}(R+\rho) 
-2(R+\rho)R\biggl(1-T^2_{,R}(R)\biggr) \cdot \nonumber \\ &  \cdot 
T_{,R}(R+\rho)\pi_{\psi}(R+\rho)  
\psi_{,R}(R+\rho) +\frac{1}{2}(R+\rho)^2R  
\biggl(1-T^2_{,R}(R)\biggr)\cdot \nonumber \\ & \cdot 
\psi^2_{,R}(R+\rho)\biggr]\biggr\}\biggr]
\biggr\} |\Psi(0,\psi(R+\rho))\!>\biggr|^2
  \nonumber
\end{align} 
In order to be able to calculate the last expression we exploit the fact that
the shift $\rho$ is very small so one may expand $\psi(R+\rho)$,
$T(R+\rho)$,  $\psi^2_{,R}(R+\rho)$, $T^2_{,R}(R+\rho)$    in a Taylor series 
and keeping the first two terms as follows  

\begin{align}  & \psi(R+\rho)=
\psi(R)+\rho\frac{d\psi(R)}{dR} \nonumber \\ & 
T(R+\rho)=T(R)+\rho\frac{dT(R)}{dR}  \tag{$B_3$} \label{$B_3$} \\ & 
\psi^2_{,R}(R+\rho)=\bigl(\psi_{,R}(R)+\rho(\frac{d\psi(R)}{dR})_{,R}\bigr)^2
=(\psi_{,R}(R))^2+ \nonumber \\ & + \rho^2(\frac{d^2\psi(R)}{d^2R})^2+ 
 2
\rho\psi(R)_{,R}\frac{d^2\psi(R)}{dR^2}
\nonumber \\ & 
T^2_{,R}(R+\rho)=\bigl(T_{,R}(R)+\rho(\frac{dT(R)}{dR})_{,R}\bigr)^2
=(T_{,R}(R))^2+\nonumber \\ & + \rho^2(\frac{d^2T(R)}{dR^2})^2+  2
\rho T(R)_{,R}\frac{d^2T(R)}{dR^2} 
\nonumber  \end{align}

Note that $\pi_{\psi}(R+\rho)$ and its square $\pi^2_{\psi}(R+\rho)$, 
 given by the second  of Eqs (\ref{e47}),  are
variational derivatives operators   
that operate upon an arbitrary function
of the shifted state 
$\psi(R+\rho)$, i.e., $f(\psi(R+\rho))$ so it is obvious 
that operating  with either 
$\pi_{\psi}(R+\rho)$ or  $\pi^2_{\psi}(R+\rho)$ upon 
$f(\psi(R+\rho))$ is
identical with respectively operating with either $\pi_{\psi}(R)$ or 
 $\pi^2_{\psi}(R)$ upon it. That is, using the second of Eq (\ref{e47}), 
  one have 
\begin{align} & \pi_{\psi}(R+\rho)f(\psi(R+\rho))
=-i\frac{\delta (f(\psi(R+\rho)))}{\delta (\psi(R+\rho))}= \nonumber \\ & =
-i\frac{\delta (f(\psi(R+\rho)))}{\delta (\psi(R))} 
=\pi_{\psi}(R)f(\psi(R+\rho)) \tag{$B_4$} \label{$B_4$} \\ &   
 \pi^2_{\psi}(R+\rho)f(\psi(R+\rho))
=-\frac{\delta^2 (f(\psi(R+\rho)))}{\delta^2 (\psi(R+\rho))}= \nonumber \\ & =
-\frac{\delta^2 (f(\psi(R+\rho)))}{\delta^2 (\psi(R))} = 
\pi^2_{\psi}(R)f(\psi(R+\rho)) \nonumber 
\end{align}

Using Eqs (\ref{$B_3$})-(\ref{$B_4$}) one may write the probability from Eq
(\ref{$B_2$}) as
\begin{align}  &  Pr^{(1)}(|\Psi(0,\psi(R)|\!>,\rho)= \biggl|
<\!\Psi(0,\psi(R))|
\exp\biggl\{\rho\biggl[
\frac{i\biggl(A\pi^2_{\psi}(R)+B\pi_{\psi}(R)+C\biggr)}{D}\biggr]\biggr\}
\nonumber \\ &  
|\Psi(0,\psi(R+\rho)|\!>\biggr|^2,  \tag{$B_5$} \label{$B_5$} 
\end{align}
where $A$, $B$, $C$  and $D$ are given by 

\begin{align} &
A=\bigl(1-T^2_{,R}(R)\bigr)\bigl(T(R)-\frac{dT(R)}{dR}\bigr)-
T(R)(R+\rho)\biggl(\rho(\frac{d^2T(R)}{dR^2})^2+ \nonumber \\ & + 2
T_{,R}(R)\frac{d^2T(R)}{dR^2}\biggr)
 \nonumber \\ & 
B=2R(R+\rho)\biggl\{T_{,R}(R)\psi_{,R}(R)T(R)
\biggl(\rho(\frac{d^2T(R)}{dR^2})^2+2
 T_{,R}(R)\frac{d^2T(R)}{dR^2}\biggr)
+ \nonumber \\ & +
\biggl(1-T^2_{,R}(R)\biggr)\biggl\{\biggl(T(R)+\rho\frac{dT(R)}{dR}\biggr)
\biggl[\frac{d^2T(R)}{dR^2}
\biggl(\psi_{,R}(R)+\rho\frac{d^2\psi(R)}{dR^2}\biggr)+ \nonumber \\ & 
+T_{,R}(R)\frac{d^2\psi(R)}{dR^2}\biggr]+
\rho T_{,R}(R)\frac{dT(R)}{dR}\frac{d^2\psi(R)}{dR^2}\biggr\}\biggr\}
  \tag{$B_6$}
\label{$B_6$} \\ 
&
C=-\frac{R^2}{2}(R+\rho)\psi^2_{,R}(R)T(R)
\biggl(\rho(\frac{d^2T(R)}{dR^2})^2+2 
T_{,R}(R)\frac{d^2T(R)}{dR^2}\biggr) - \nonumber \\ & 
-\frac{R}{2}(R+\rho)^2\biggl(1-T^2_{,R}(R)\biggr) 
\biggl(T(R)+\rho\frac{dT(R)}{dR}\biggr)
\biggl(\rho(\frac{d^2\psi(R)}{dR^2})^2  +\nonumber \\ & 
+2\psi_{,R}(R)\frac{d^2\psi(R)}{dR^2}\biggr)
-  
\frac{R}{2}(R+\rho)\biggl(1-T^2_{,R}(R)\biggr)
\psi^2_{,R}(R)\biggl(T(R)+ \nonumber \\ & + (R+\rho)\frac{dT(R)}{dR}\biggr) 
\nonumber \\ &
 D=2R(R+\rho)
\biggl(1-T^2_{,R}(R)\biggr)\biggl(1-T^2_{,R}(R+\rho)\biggr) \nonumber   
\end{align}

In order to continue we must overcome the commutation problem resulting from the
presence of $\pi_{\psi}(R)$, $\pi^2_{\psi}(R)$,   $<\!\Psi(0,\psi(R))|$ and 
$|\Psi(0,\psi(R+\rho))\!>$ in Eq (\ref{$B_5$})  (see the second of 
Eqs (\ref{e47})). For this purpose one may begin from expanding the exponential
functions of the probability expression Eq (\ref{$B_5$}) in Taylor series and 
then 
use the commutation relations between the resulting 
integral powers of $\pi_{\psi}$ and $|\Psi(0,\psi(R))\!>$ which may be derived
by using the second of Eqs (\ref{e47}). Thus,  one may, for example,  see that the following
respective 
commutation relations between $|\Psi(0,\psi(R))\!>$ and $\pi_{\psi}(R)$, 
$\pi^2_{\psi}(R)$,   
$\pi^3_{\psi}(R)$ 
hold

\begin{align} & [|\Psi(0,\psi(R))\!>,\pi_{\psi}(R)]=
i\frac{d(|\Psi(0,\psi(R))\!>)}{d(\psi(R))} \nonumber \\ & 
[|\Psi(0,\psi(R))\!>,\pi^2_{\psi}(R)]=
2i\frac{d(|\Psi(0,\psi(R))\!>)}{d(\psi(R))}\pi_{\psi}+
\frac{d^2(|\Psi(0,\psi(R))\!>)}{d^2(\psi(R))} \tag{$B_7$} \label{$B_7$} \\ & 
[|\Psi(0,\psi(R))\!>,\pi^3_{\psi}(R)]=
3i\frac{d(|\Psi(0,\psi(R))\!>)}{d(\psi(R))}\pi^2_{\psi}+
\frac{d^2(|\Psi(0,\psi(R))\!>)}{d^2(\psi(R))}\pi_{\psi}(R) \nonumber 
\end{align} 
Note that although we refer to $|\Psi(0,\psi(R))\!>$ as a state we should remember that it is actually the solution (\ref{e50}) of the Schroedinger-type equation (\ref{e48}) and  is a function of the operators $T(R)$, $\psi(R)$  so one may calculate the commutation between it and the operator  $\pi_{\psi}(R)$ (see similar relations at P. 177-178 in \cite{schiff}). 
Thus,  using mathematical induction and the last relations one may realize 
that for any integral value $n$  the following 
commutation relation is obtained
\begin{align}  &   [|\Psi(0,\psi(R))\!>,\pi^n_{\psi}(R)]=
ni\frac{d(|\Psi(0,\psi(R))\!>)}{d(\psi(R))}\pi^{(n-1)}_{\psi}(R)+
 \tag{$B_8$} \label{$B_8$} \\  & + 
\frac{d^2(|\Psi(0,\psi(R))\!>)}{d^2(\psi(R))}\pi^{(n-2)}_{\psi}(R) \nonumber 
\end{align}
 Note that the commutation relations in Eqs (\ref{$B_7$})-(\ref{$B_8$}) hold for
 any general function of $\psi(R)$ and not only for $|\Psi(0,\psi(R))\!>$.  
 Using the last equations one may find the appropriate expressions for
  operating with the operators  
  $\exp(i\rho\frac{A\pi^2_{\psi}(R)}{E})$ and  
  $\exp(i\rho\frac{B\pi_{\psi}(R)}{E})$  upon $|\Psi(0,\psi(R))\!>$
  as 
  \begin{align} &
  \exp(i\rho\frac{A}{E}\pi^2_{\psi}(R))|\Psi(0,\psi(R))\!>=
  \biggl(1+i\rho\frac{A}{E}\pi^2_{\psi}(R)-\rho^2\frac{A^2}{2!E^2}\pi^4_{\psi}(R)
 - \nonumber \\ & - i\rho^3\frac{A^3}{3!E^3}\pi^6_{\psi}(R)+\ldots\biggr)
  |\Psi(0,\psi(R))\!>=
  |\Psi(0,\psi(R))\!>+i\rho\frac{A}{E}
 \biggl(|\Psi(0,\psi(R))\!>\pi^2_{\psi}- \nonumber \\ & - 
2i\frac{d(|\Psi(0,\psi(R))\!>)}{d(\psi(R))}\pi_{\psi}(R)- 
\frac{d^2(|\Psi(0,\psi(R))\!>)}{d^2(\psi(R))}\biggr) -
\rho^2\frac{A^2}{2!E^2}\biggl(|\Psi(0,\psi(R))\!> \nonumber \\ & \cdot 
\pi^4_{\psi}(R) - 
4i\frac{d(|\Psi(0,\psi(R))\!>)}{d(\psi(R))}\pi^3_{\psi}(R)- 
\frac{d^2(|\Psi(0,\psi(R))\!>)}{d^2(\psi(R))}\pi^2_{\psi}\biggr)+
\ldots = \tag{$B_9$} \label{$B_9$} \\ & = 
|\Psi(0,\psi(R))\!>\exp(i\rho\frac{A}{E}\pi^2_{\psi}(R))-
\frac{1}{\pi^2_{\psi}(R)}\frac{\delta^2(|\Psi(0,\psi(R))\!>)}{\delta^2(\psi(R)} 
 \exp(i\rho\frac{A}{E}\pi^2_{\psi}(R))+ \nonumber \\ & + 
\frac{1}{\pi^2_{\psi}(R)}\frac{\delta^2(|\Psi(0,\psi(R))\!>)}{\delta^2(\psi(R)}+
2\rho\frac{A}{E}\pi_{\psi}(R)\frac{\delta(|\Psi(0,\psi(R))\!>)}{\delta(\psi(R))}
\exp(i\rho\frac{A}{E}\pi^2_{\psi}(R)) 
\nonumber \end{align}

 \begin{align} & 
  \exp(i\rho\frac{B}{E}\pi_{\psi}(R))|\Psi(0,\psi(R))\!>=
  \biggl(1+i\rho\frac{B}{E}\pi_{\psi}(R)-\rho^2\frac{B^2}{2!E^2}\pi^2_{\psi}(R)
 - \nonumber \\ & - 
  i\rho^3\frac{B^3}{3!E^3}\pi^3_{\psi}(R)+\ldots\biggr)  |\Psi(0,\psi(R))\!>=
  |\Psi(0,\psi(R))\!>+i\rho\frac{B}{E}
 \biggl(|\Psi(0,\psi(R))\!>\pi_{\psi}- \nonumber \\ & - 
i\frac{d(|\Psi(0,\psi(R))\!>)}{d(\psi(R))}\biggr) -
\rho^2\frac{B^2}{2!E^2}\biggl(|\Psi(0,\psi(R))\!>\pi^2_{\psi}(R)-
2i\frac{d(|\Psi(0,\psi(R))\!>)}{d(\psi(R))}\pi_{\psi}- \nonumber \\  & - 
\frac{d^2(|\Psi(0,\psi(R))\!>)}{d^2(\psi(R))}\biggr)+
\ldots  =
|\Psi(0,\psi(R))\!>\exp\biggl(i\rho\frac{B}{E}\pi_{\psi}(R)\biggr)- 
\tag{$B_{10}$} 
\label{$B_{10}$} \\
& - 
\frac{1}{\pi^2_{\psi}(R)}\frac{\delta^2(|\Psi(0,\psi(R))\!>)}{\delta^2(\psi(R)} 
 \cdot \exp(i\rho\frac{B}{E}\pi_{\psi}(R))+  
\frac{1}{\pi^2_{\psi}(R)}\frac{\delta^2(|\Psi(0,\psi(R))\!>)}{\delta^2(\psi(R)}+
 \nonumber \\ & + i\rho\frac{B}{E\pi_{\psi}}\frac{\delta^2(|\Psi(0,\psi(R))\!>)}{\delta^2(\psi(R)}+ \rho\frac{B}{E}\frac{\delta(|\Psi(0,\psi(R))\!>)}{\delta(\psi(R))}
\exp(i\rho\frac{B}{E}\pi_{\psi}(R)) 
\nonumber \end{align}

Using Eqs (\ref{$B_9$})-(\ref{$B_{10}$}) one may obtain for the 
combined operation \\ 
$\exp\biggl(i\rho\frac{A}{E}\pi^2_{\psi}(R)\biggr)
\exp\biggl(i\rho\frac{B}{E}\pi_{\psi}(R)\biggr)$ upon $|\Psi(0,\psi(R))\!>$ 
\begin{align} & \exp\biggl(i\rho\frac{A}{E}\pi^2_{\psi}(R)\biggr)
\exp\biggl(i\rho\frac{B}{E}\pi_{\psi}(R)\biggr)|\Psi(0,\psi(R)\!>=
\exp\biggl(i\rho\frac{A\pi^2_{\psi}(R)}{E}\biggr)
\biggl\{|\Psi(0,\psi(R)\!> \cdot \nonumber \\ & \cdot 
\exp\biggl(i\rho\frac{B}{E}\pi_{\psi}(R)\biggr)-
\frac{1}{\pi^2_{\psi}(R)}\frac{\delta^2(|\Psi(0,\psi(R))\!>)}{\delta^2(\psi(R)} 
 \cdot \exp\biggl(i\rho\frac{B}{E}\pi_{\psi}(R)\biggr)+ \nonumber \\ & + 
\frac{1}{\pi^2_{\psi}(R)}\frac{\delta^2(|\Psi(0,\psi(R))\!>)}
{\delta^2(\psi(R)}+ 
i\rho\frac{B}{E\pi_{\psi}}\frac{\delta^2(\Psi(0,\psi(R))|\!>)}{\delta^2(\psi(R))}+
\rho\frac{B}{E}\frac{\delta(|\Psi(0,\psi(R))\!>)}{\delta(\psi(R))} \cdot
\nonumber \\ & \cdot
\exp(i\rho\frac{B}{E}\pi_{\psi}(R))\biggr\}=  
|\Psi(0,\psi(R))\!>\exp\biggl(i\rho\frac{A}{E}\pi^2_{\psi}(R)\biggr) 
\exp\biggl(i\rho\frac{B}{E}\pi_{\psi}(R)\biggr)+ \nonumber \\ & + 
\frac{1}{\pi^2_{\psi}(R)}\frac{\delta^2(\Psi(0,\psi(R))|\!>)}
{\delta^2(\psi(R))}
  \biggl[\exp\biggl(i\rho
 \frac{B}{E}\pi_{\psi}(R)\biggr) 
+\exp\biggl(i\rho\frac{A}{E}\pi^2_{\psi}(R)\biggr) 
 - \tag{$B_{11}$} \label{$B_{11}$} \\ & - 
 2\exp\biggl(i\rho\frac{A}{E}\pi^2_{\psi}(R)\biggr)  
 \exp\biggl(i\rho\frac{B}{E}\pi_{\psi}(R)\biggr) \biggr]+ 
\rho\frac{\biggl(2A\pi_{\psi}(R)+B\biggr)}{E}\frac{\delta(|\Psi(0,\psi(R))\!>)}
{\delta(\psi(R))} \cdot \nonumber \\ & \cdot 
\exp\biggl(i\rho\frac{A}{E}\pi^2_{\psi}(R)\biggr)   
\exp\biggl(i\rho\frac{B}{E}\pi_{\psi}(R)\biggr)+ 
i\rho\frac{B}{E\pi_{\psi}}\frac{\delta^2(|\Psi(0,\psi(R))\!>)}
{\delta^2(\psi(R)}\exp\biggl(i\rho\frac{A}{E}\pi^2_{\psi}(R)\biggr) 
\nonumber \end{align} 

Substituting from  the last  equations (\ref{$B_9$})-(\ref{$B_{11}$}) into  
Eq (\ref{$B_5$}) for
the probability one obtains
 
 \begin{align} & Pr^{(1)}(|\Psi(0,\psi(R))\!>,\rho)=\biggl| <\!\Psi(0,\psi(R)|)
\exp\biggl\{\rho\biggl[
\frac{i\biggl(A\pi^2_{\psi}(R)+B\pi_{\psi}(R)+C\biggr)}{D}\biggr]\biggr\} 
\cdot \nonumber \\ & \cdot |\Psi(0,\psi(R+\rho)\!>\biggr|^2= 
\biggl|<\!\Psi(0,\psi(R)|)
\exp\biggl(i\rho\frac{C}{D}\biggr) \cdot 
\exp\biggl(i\rho\frac{A}{D}\pi^2_{\psi}(R)\biggr)\cdot \nonumber \\ & \cdot 
\exp\biggl(i\rho\frac{B}{D}\pi_{\psi}(R)\biggr) 
 |\Psi(0,\psi(R+\rho)\!>\biggr|^2 
 =\biggl| <\!\Psi(0,\psi(R))\exp\biggl(i\rho\frac{C}{D}\biggr)\cdot 
 \tag{$B_{12}$} 
\label{$B_{12}$} \\ & \cdot 
\biggl\{|\Psi(0,\psi(R+\rho))\!>
\exp\biggl(i\rho\frac{A}{D}\pi^2_{\psi}(R)\biggr)  
\exp\biggl(i\rho\frac{B}{D}\pi_{\psi}(R)\biggr)+ 
\frac{1}{\pi^2_{\psi}(R)} \nonumber \\ & 
\cdot \frac{\delta^2(|\Psi(0,\psi(R+\rho))\!>)}
{\delta^2(\psi(R))} 
  \biggl[\exp\biggl(i\rho
 \frac{B}{E}\pi_{\psi}(R)\biggr) 
+\exp\biggl(i\rho\frac{A}{D}\pi^2_{\psi}(R)\biggr) - \nonumber \\ & 
 -2\exp\biggl(i\rho\frac{A}{D}\pi^2_{\psi}(R)\biggr) 
 \exp\biggl(i\rho\frac{B}{D}\pi_{\psi}(R)\biggr) \biggr]+
\rho\frac{\biggl(2A\pi_{\psi}(R)+B\biggr)}{D} \nonumber \\ & \cdot 
\frac{\delta(|\Psi(0,\psi(R+\rho))\!>)}
{\delta(\psi(R))} 
\exp\biggl(i\rho\frac{A}{D}\pi^2_{\psi}(R)\biggr) 
\exp\biggl(i\rho\frac{B}{D}\pi_{\psi}(R)\biggr)+ \nonumber \\ & +  
i\rho\frac{B}{E\pi_{\psi}}\frac{\delta^2(|\Psi(0,\psi(R+\rho))\!>)}
{\delta^2(\psi(R)}\exp\biggl(i\rho\frac{A}{D}\pi^2_{\psi}(R)\biggr)\biggr\}
\biggr|^2
\nonumber
\end{align}
Note that since, generally, different powers of the same operator commute with each other we have equated the exponent of the sum of $\pi^2_{\psi}(R)$ and $\pi_{\psi}(R)$ to the product of the exponents of the corresponding operators. Now,  
in order to simplify the following calculation we  
 denote the  two complex
factors  in Eq (\ref{$B_{12}$}) which respectively multiply 
 $\frac{\delta^2(\Psi(0,\psi(R+\rho))|\!>)}
{\delta^2(\psi(R))}$ and $\frac{\delta(\Psi(0,\psi(R+\rho))|\!>)}
{\delta(\psi(R))}$ by $F(\rho)$ and $G(\rho)$ and use the second of Eqs 
(\ref{e47}) 
to write these variational
derivatives as $\frac{\delta(|\Psi(0,\psi(R+\rho))\!>)}
{\delta(\psi(R))}=i\pi_{\psi}(R)\Psi(0,\psi(R+\rho))\!>$ and 
$\frac{\delta^2(|\Psi(0,\psi(R+\rho))\!>)}
{\delta^2(\psi(R))}=-\pi^2_{\psi}(R)\Psi(0,\psi(R+\rho))\!>$.  
Thus, one may write
the two expressions in Eq (\ref{$B_{12}$}) related to these variational derivatives as  
\begin{align} &  
\frac{1}{\pi^2_{\psi}(R)}\frac{\delta^2(|\Psi(0,\psi(R+\rho))\!>)}
{\delta^2(\psi(R))} 
\biggl[\exp\biggl(i\rho\frac{B}{D}\pi_{\psi}(R)\biggr) +
 \exp\biggl(i\rho
\frac{A}{D}\pi^2_{\psi}(R)\biggr)\biggl(1
+i\rho\frac{B}{D}\pi_{\psi}(R)\biggr) - \nonumber \\ & 
   - 2\exp\biggl(i\rho\frac{A}{D}\pi^2_{\psi}(R)\biggr) 
 \exp\biggl(i\rho\frac{B}{D}\pi_{\psi}(R)\biggr)\biggr] =
 -\pi^2_{\psi}(R)|\Psi(0,\psi(R+\rho))\!>F(\rho)  
  \tag{$B_{13}$} \label{$B_{13}$} \\ & 
\rho\frac{\biggl(2A\pi_{\psi}(R)+B\biggr)}{D}\frac{\delta(|\Psi(0,\psi(R+\rho))\!>)}
{\delta(\psi(R))}  
\exp\biggl(i\rho\frac{A}{D}\pi^2_{\psi}(R)\biggr)
\exp\biggl(i\rho\frac{B}{D}\pi_{\psi}(R)\biggr)= \nonumber \\ & =
 i\pi_{\psi}(R)|\Psi(0,\psi(R+\rho))\!>G(\rho)
\nonumber \end{align}
Note that we discuss here a half-parametrized formalism and 
 a representation in which the canonical coordinates
$T(R)$ and $\psi(R)$ are diagonal (see discussion after Eq (\ref{e47})) so the
state functional depends only upon these two cordinates i.e.,
 $\Psi(T(R),\psi(R))$. This representation enables one to introduces the Dirac
 constraint as the Schroedinger-type form of Eq (\ref{e48}) and also to write the
 coresponding eigenvalue constraint $\pi_{\psi}(R)|\Psi(0,\psi(R+\rho))\!>=
 \eta|\Psi(0,\psi(R+\rho))\!> $ where $\eta$ is an eigenvalue related to the
 operator $\pi_{\psi}(R)$. Note that a similar momentum eigenvalue equation is discussed in \cite{kuchar1} (see Eq (64) in \cite{kuchar1}). Thus, using the last eigenvalue relation, the
 normalization \cite{kuchar1} of the state functionals $\Psi(0,\psi(R))$  and
 the complex character  of the expressions $F(\rho)$ and $G(\rho)$ from Eq
 (\ref{$B_{13}$}) which enables one to represent them in terms of their real and
 imaginary parts 
one may write the former probability from Eq (\ref{$B_{12}$}) as
 \begin{align} & Pr^{(1)}(|\Psi(0,\psi(R))\!>,\rho)=
\biggl| \exp(i\rho\frac{C}{D})<\!\Psi(0,\psi(R))|
\biggl\{|\Psi(0,\psi(R+\rho))\!> \cdot  \nonumber \\ & \cdot 
\exp\biggl(i\frac{\rho}{D}
\biggl(A\pi^2_{\psi}(R) + B\pi_{\psi}(R)\biggr)\biggr)+ 
i
\pi_{\psi}(R)|\Psi(0,\psi(R+\rho\!>G(\rho)- \tag{$B_{14}$} 
\label{$B_{14}$} \\ & - \pi^2_{\psi}(R)
|\Psi(0,\psi(R+\rho))\!>F(\rho) \biggr\}\biggr|^2=
 \biggl|<\!\Psi(0,\psi(R))|\Psi(0,\psi(R+\rho))\!>
\exp(i\rho\frac{C}{D})\biggr|^2 \cdot  \nonumber \\ & 
\cdot \biggl|\biggr\{\exp\biggl[i\frac{\rho}{D}
\biggl(A\pi^2_{\psi}(R) 
+ B\pi_{\psi}(R)\biggr)\biggr]+  
i\eta \biggl(\Re G(\rho)+i\Im G(\rho)\biggr)-\eta^2\biggl( \Re F(\rho)+
i\Im F(\rho)\biggr)\biggr\}\biggr|^2 
\nonumber
\end{align}
 The real and imaginary parts of $G(\rho)$ and $F(\rho)$ 
denoted, respectively,  by  $\Re G(\rho)$, 
 $\Re F(\rho)$, $\Im G(\rho)$ and $\Im F(\rho)$,   are given by Eq 
 (\ref{$B_{13}$}). Thus, substituting from Eq (\ref{$B_{13}$}) into 
 Eq (\ref{$B_{14}$}) and using the following  trigonometric product identities; 
 $\sin(x)\sin(y)=\frac{1}{2}\biggl(\cos(x-y)-\cos(x+y)\biggr)$,  
 $\cos(x)\cos(y)=\frac{1}{2}\biggl(\cos(x-y)+\cos(x+y)\biggr)$, 
  $\sin(x)\cos(y)=\frac{1}{2}\biggl(\sin(x-y)+\sin(x+y)\biggr)$ 
  and  $\sin^2(x)+\cos^2(y)=1$ one obtains for
  the probability $Pr^{(1)}(|\Psi(0,\psi(R))\!>,\rho)$
   \begin{align} &
  Pr^{(1)}(|\Psi(0,\psi(R))\!>,\rho)=1+4\eta^2+6\eta^4+
\rho^2\eta^4(\frac{B\pi_{\psi}(R)}{D})^2+
 \eta^2\frac{\rho^2}{D^2}\biggl(2A\pi_{\psi}(R)+ \nonumber \\ & + 
 B\biggr)^2+ 
 \sin(\frac{\rho}{D}B\pi_{\psi}(R))
 \biggl\{2\eta^3\frac{\rho}{D\pi^2_{\psi}(R)}\biggl(2A\pi_{\psi}(R)+B\biggr)
 -2\eta^2\frac{\rho B\pi_{\psi}(R)}{D}\bigl(1+ \nonumber \\ & + 
 2\eta^2\bigr)\biggr\}-\cos(\frac{\rho}{D}B\pi_{\psi}(R))
 \biggl\{2\eta^3\frac{\rho^2B}{D^2\pi_{\psi}(R)}\biggl(2A\pi_{\psi}(R)+B\biggr)
 +2\eta^2\bigl(1+2\eta^2\bigr)\biggr\}+ \tag{$B_{15}$} \label{$B_{15}$} \\ & + 
2\eta^3\frac{\rho}{D\pi^2_{\psi}(R)}
\biggl(2A\pi_{\psi}(R)+B\biggr) \sin(\frac{\rho}{D}A\pi^2_{\psi}(R))
-2\eta^2\bigl(1+2\eta^2\bigr)\cos(\frac{\rho}{D}A\pi^2_{\psi}(R))- \nonumber \\
& -
2\eta^4\frac{\rho B\pi_{\psi}(R)}{D}
\sin\biggl(\frac{\rho}{D}\biggl(A\pi^2_{\psi}(R)-B\pi_{\psi}(R)\biggr)\biggr)
+2\eta^4\cos\biggl(\frac{\rho}{D}\biggl(A\pi^2_{\psi}(R)-
B\pi_{\psi}(R)\biggr)\biggr)
 \nonumber \end{align}

\markright{REFERENCES}
 
  \bigskip \bibliographystyle{plain}

\begin{thebibliography}{99}
\bibitem{note1}   The 
 problem of quantization is found in the literature to refer  to gravitational
field \cite{schwinger1} and also to gravitational wave \cite{kuchar1}.  
Since we use here some results from \cite{kuchar1} we also adopt the
terminology there of gravitational wave.
\end{thebibliography}

\end{document}